\documentclass[twocolumn,twocolappendix]{aastex631}

\graphicspath{{./}{figures/}} 

\newcommand\sersic{S\'ersic}

\newcommand\gampen{GaMPEN}
\newcommand\lenstronomy{LENSTRONOMY}
\newcommand\gb{\textit{g}}
\newcommand\rb{\textit{r}}
\newcommand\ib{\textit{i}}

\shorttitle{HSC Wide Morphologies}
\shortauthors{Ghosh et al.}

\usepackage{soul}
\usepackage{enumitem}
\usepackage{bm}
\usepackage{gensymb}
\usepackage{subfigure}
\usepackage{multirow}
\usepackage{array}
\usepackage{hhline}
\usepackage[frozencache,cachedir=minted-cache]{minted}
\usepackage{listings}
\PassOptionsToPackage{hyphens}{url}
\usepackage{hyperref}

\usepackage{xcolor} 
\definecolor{LightGray}{gray}{0.95}

\font\bngxi=bang10 scaled 1100

\def\*#1*#2{o\null{#2}{#1}}

\def\d#1{\oalign{\smash{#1}\crcr\hidewidth{$\!$\rm.}\hidewidth}}

\def\sh#1{\setbox0=\hbox{#1}%
     \kern-.02em\copy0\kern-\wd0
     \kern.04em\copy0\kern-\wd0
     \kern-.02em\raise.0433em\box0 }
\usepackage{CJK}

\begin{document}
\begin{CJK*}{UTF8}{gbsn}

\title{Morphological Parameters and Associated Uncertainties for 8 Million Galaxies in the Hyper Suprime-Cam Wide Survey}

\author[0000-0002-2525-9647]{Aritra Ghosh ({\bngxi Airt/r \*gh*eaSh})}
\affil{Department of Astronomy, Yale University, New Haven, CT, USA}
\affil{Yale Center for Astronomy and Astrophysics, New Haven, CT, USA}
\email{aritra.ghosh@yale.edu; aritraghsh09@gmail.com}

\author[0000-0002-0745-9792]{C. Megan Urry}
\affil{Yale Center for Astronomy and Astrophysics, New Haven, CT, USA}
\affil{Department of Physics, Yale University, New Haven, CT, USA}

\author[0000-0003-1164-0268]{Aayush Mishra}
\affil{Department of Physics, Indian Institute of Science Education and Research, Bhopal, India}

\author[0000-0003-3544-3939]{Laurence Perreault-Levasseur}
\affil{Department of Physics, Univesit\'e de Montr\'eal, Montr\'eal, Canada}
\affiliation{Ciela Institute, Montr\'eal, Canada}
\affiliation{Mila - Quebec Artificial Intelligence Institute, Montr\'eal, Canada}
\affil{Center for Computational Astrophysics, Flatiron Institute, New York, NY, USA}

\author[0000-0002-5554-8896]{Priyamvada Natarajan}
\affil{Department of Astronomy, Yale University, New Haven, CT, USA}
\affil{Department of Physics, Yale University, New Haven, CT, USA}
\affil{Black Hole Initiative, Harvard University, Cambridge, MA, USA}

\author[0000-0002-1233-9998]{David B. Sanders}
\affiliation{
Institute for Astronomy, University of Hawai`i, Honolulu, HI, USA}

\author[0000-0002-6766-5942]{Daisuke Nagai}
\affil{Department of Astronomy, Yale University, New Haven, CT, USA}
\affil{Yale Center for Astronomy and Astrophysics, New Haven, CT, USA}
\affil{Department of Physics, Yale University, New Haven, CT, USA}

\author[0000-0003-4056-7071]{Chuan Tian (田川)}
\affil{Yale Center for Astronomy and Astrophysics, New Haven, CT, USA}
\affil{Department of Physics, Yale University, New Haven, CT, USA}

\author[0000-0002-1697-186X]{Nico Cappelluti}
\affiliation{Department of Physics, University of Miami, Coral Gables, FL, USA}

\author[0000-0001-9187-3605]{Jeyhan S. Kartaltepe}
\affil{Laboratory for Multiwavelength Astrophysics, School of Physics and Astronomy,\protect\\ Rochester Institute of Technology, Rochester, NY, USA}

\author[0000-0003-2284-8603]{Meredith C. Powell}
\affil{Kavli Institute of Particle Astrophysics and Cosmology, Stanford University, Stanford, CA, USA}

\author{Amrit Rau}
\affil{Department of Computer Science, Yale University, New Haven, CT, USA}

\author[0000-0001-7568-6412]{Ezequiel Treister}
\affiliation{Instituto de Astrof\'isica, Facultad de F\'isica, Pontificia Universidad Cat\'olica de Chile, Santiago, Chile}

\begin{abstract}
We use the Galaxy Morphology Posterior Estimation Network (GaMPEN) to estimate morphological parameters and associated uncertainties for $\sim 8$ million galaxies in the Hyper Suprime-Cam (HSC) Wide survey with $z \leq 0.75$ and $m \leq 23$. GaMPEN is a machine learning framework that estimates Bayesian posteriors for a galaxy's bulge-to-total light ratio ($L_B/L_T$), effective radius ($R_e$), and flux ($F$). By first training on simulations of galaxies and then applying transfer learning using real data, we trained GaMPEN with $<1\%$ of our data-set. This two-step process will be critical for applying machine learning algorithms to future large imaging surveys, such as the Rubin-Legacy 
Survey of Space and Time (LSST), the Nancy Grace Roman Space Telescope (NGRST), and Euclid. By comparing our results to those obtained using light-profile fitting, we demonstrate that GaMPEN's predicted posterior distributions are well-calibrated ($\lesssim 5\%$ deviation) and accurate. This represents a significant improvement over light profile fitting algorithms which underestimate uncertainties by as much as $\sim60\%$. For an overlapping sub-sample, we also compare the derived morphological parameters with values in two external catalogs and find that the results agree within the limits of uncertainties predicted by GaMPEN. This step also permits us to define an empirical relationship between the S\'ersic index and $L_B/L_T$ that can be used to convert between these two parameters. The catalog presented here represents a significant improvement in size ($\sim10 \times $), depth ($\sim4$ magnitudes), and uncertainty quantification over previous state-of-the-art bulge+disk decomposition catalogs. With this work, we also release GaMPEN's source code and trained models, which can be adapted to other data sets.
\end{abstract}


\section{Introduction} \label{sec:intro}

The morphology of galaxies has been shown to be related to various other fundamental properties of galaxies and their environment, including galaxy mass, star formation rate, stellar kinematics, merger history, cosmic environment, and the influence of supermassive black holes \citep[e.g.,][]{Bender1992DynamicallyProperties,Tremaine2002TheCorrelation,pozzetti_10, wuyts_11, Huertas-Company2016MassCANDELS,powell_17, shimakawa_2021, Dimauro2022CoincidenceGrowth}. Therefore, quantitative measures of the morphological parameters for large samples of galaxies at different redshifts are of fundamental importance in understanding the physics of galaxy formation and evolution. 

Distributions of morphological quantities alone can place powerful constraints on possible galaxy formation scenarios. And when combined with other physical quantities, they can provide key insights into evolutionary processes at play or even reveal the role of new physical  mechanisms that impact evolution \citep[e.g.,][]{Kauffmann2004TheGalaxies,Weinmann2006PropertiesMass,Schawinski2007TheGalaxies,vanderWel2008TheMass,Schawinski2014TheGalaxies}. However, such studies often involve subtle correlations or hidden variables within strong correlations that demand greater statistics and measurement precision than what has been available in the preceding decades. 
An often-overlooked factor in such studies has been the computation of robust uncertainties. The computation of full Bayesian posteriors for different morphological parameters is crucial for drawing scientific inferences that account for uncertainty and, thus are indispensable in 
the derivation of robust scaling relations  \citep[e.g.,][]{Bernardi2013TheProfile, vanderWel20143D-HST+CANDELS:3} or tests of theoretical models using morphology \citep[e.g.,][]{Schawinski2014TheGalaxies}.

A quantitative description of galaxy morphology is typically expressed in terms of its structural parameters --- brightness, shape, and size --- all of which can be determined by fitting a single two dimensional analytic light profile to the galaxy image \citep[e.g.,][]{vdw_12,tarsitano_18}. However, moving beyond single-component determinations by using separate components to analyze galaxy sub-structure (e.g., disk, bulge, bar, etc.) can provide us additional insights into the formation mechanisms of these components: bulges, disks, and bars may be formed as a result of secular evolution \citep[e.g.,][]{kormendy_1979,kormendy_2004, genzel_2008, sellwood_2014} or due to the interaction of disk instabilities with smooth and clumpy cold streams \citep[e.g.,][]{dekel_09a,dekel_09b}. In this sense, contrary to what is often expected, bulges can also be formed without major galaxy mergers. 

Over the last decade, machine learning (ML) has been increasingly used for a wide variety of tasks---from identifying exoplanets to studying black holes \citep[e.g.,][]{ml_pz,ml_sz,Shallue2018IdentifyingKepler-90,Sharma2020ApplicationClassification,Natarajan2021QuasarNet:Holes}. Unsurprisingly, these algorithms have become increasingly popular for determining galaxy morphology as well. \citep[e.g.,][]{Dieleman2015Rotation-invariantPrediction, Huertas-Company2015ALEARNING, Tuccillo2018DeepFitting, gamornet_paper, Hausen2020MorpheusData, Walmsley2020GalaxyLearning, Cheng2021GalaxyNetworks, Vega-Ferrero2021PushingSurvey, Tarsitano2022ImageLearning}. The use of these techniques has been driven by the fact that  traditional methods of analyzing morphologies---visual classification and template fitting---are not scalable to the data volume expected from future surveys such as the Vera Rubin Observatory Legacy Survey of Space and Time \citep[LSST;][]{lsst}, the Nancy Grace Roman Space Telescope \citep[NGRST;][]{ngrst}, and Euclid \citep{euclid}. 

Most previous applications of ML to galaxy morphology produced broad, qualitative classifications rather than numerical estimates of morphological parameters. \citet{Tuccillo2018DeepFitting} did estimate parameters of single-component \sersic{} fits. However, they did not analyze galaxy sub-structures or provide uncertainties. In order to address these challenges, in \citet{gampen_software_paper}, we introduced The Galaxy Morphology Posterior Estimation Network (\gampen{}). \gampen{} is a machine learning framework that estimates full Bayesian posteriors for a galaxy's bulge-to-total light ratio ($L_B/L_T$), effective radius ($R_e$), and flux ($F$). \gampen{} takes into account covariances between the different parameters in its predictions and has been shown to produce calibrated, accurate posteriors. 

\gampen{} can also automatically crop input galaxy cutouts to an optimal size before determining their morphology. This feature is critical given that morphology-determination ML frameworks typically require input cutouts of a fixed size, and thus, cutouts of a ``typical" size often contain secondary objects in the frame. By cropping out most secondary objects in the frame, \gampen{} can make more accurate predictions over wide ranges of redshift and magnitude. 

\begin{figure}[htb]
    \centering
    \includegraphics[width = 0.47\textwidth]{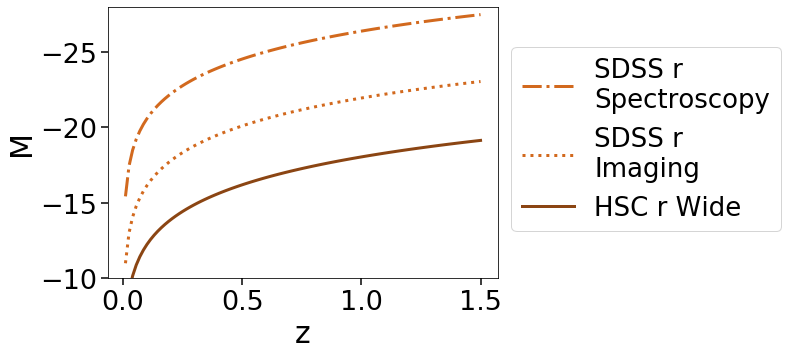}
    \caption{The limiting absolute magnitudes probed by the Hyper Suprime-Cam (HSC) Wide Survey and Sloan Digital Sky Survey (SDSS) at different redshifts.}
    \label{fig:hsc_depth}
\end{figure}

In this paper, we use \gampen{} to estimate Bayesian posteriors for $L_B/L_T$, $R_e$, and $F$ for $\sim 8$ million galaxies with $z \leq 0.75$ from the Hyper Suprime-Cam (HSC) Wide survey \citep{hsc_design}. A few recent works have studied the morphology of smaller subsets of HSC galaxies: \citet{shimakawa_2021} classified $\sim2\times10^5$ massive HSC galaxies into spiral and non-spiral galaxies, while \citet{hsc_sersic} analyzed $\sim1.5\times10^6$ HSC galaxies using single-component \sersic{} fits. To date, the state-of-the-art morphological catalog that provided bulge+disk decomposition parameters at low redshift has been that of \citet{simard_11}, which used Sloan Digital Sky Survey \citep[SDSS; ][]{sdss_tech_summary} imaging to estimate morphological parameters of $\sim1$ million $m < 18$ galaxies, most with $z < 0.2$. As Figure \ref{fig:hsc_depth} shows, HSC imaging allows us to probe much fainter magnitudes than SDSS with significantly better seeing (0.\arcsec85 for HSC-W \gb{} compared to 1.\arcsec4 for SDSS \gb{}). The catalog presented in this paper builds on \citet{simard_11} by providing an order of magnitude increase in sample size and probing four magnitudes deeper, to a higher redshift threshold. Along with estimates of parameters, this catalog also estimates robust uncertainties of the predicated parameters, which have typically been absent from previous large morphological catalogs. This catalog represents a significant step forward in our capability to quantify the shapes and sizes of galaxies in our universe.

This paper also demonstrates that ML techniques can be used to study morphology in new surveys, which do not have already-classified large training sets available. Most previous works involving ML to study galaxy morphology have depended on the availability of an extensive training set of real galaxies with known properties from the same survey or a similar one. However, if Convolutional Neural Networks (CNNs) are to replace traditional methods for morphological analysis, we must be able to use them on new surveys that do not have a morphological catalog readily available to be used as a training set. In this paper, we demonstrate that by first training on simulations of galaxies and then using real data for transfer learning (fine-tuning the simulation-trained network), we can fully train \gampen{}, while having to label $<1\%$ of our dataset. This work outlines an easy pathway to apply morphological ML techniques to upcoming large imaging surveys.  

In \S \ref{sec:data}, we describe the HSC data used in this study, along with the simulated two-dimensional light profiles. \S \ref{sec:gampen} and \S \ref{sec:method} provide a brief introduction to \gampen{} and how we train it. In \S \ref{sec:galfitting}, we outline how we determined the morphological parameters of $\sim60,000$ HSC-Wide galaxies for transfer learning and validating the results of our ML framework. \S \ref{sec:results} provides detailed information on the accuracy of \gampen{}'s predictions. \S \ref{sec:compare} compares our predictions to that of other catalogs. We end with a summary and discussion of our results in \S \ref{sec:conclusions}.

The full Bayesian posteriors for all $\sim 8$ million galaxies are being released with the publication of this work. We are also releasing the source code of \gampen{}, along with the trained models and extensive documentation and tutorials on how to use \gampen{}. The public data release is described in Appendix \ref{sec:ap:data_access}.

\section{Data} \label{sec:data}

\subsection{Hyper Suprime-Cam Data} \label{sec:hsc_data}

We apply \gampen{} to \textit{g},\textit{r},\textit{i}-band data from the Hyper Suprime-Cam (HSC) Subaru Strategic Program Public Data Release 2 \citep[PDR2;][]{hsc_pdr2}. The Subaru Strategic Program, ongoing since 2014, uses the HSC prime-focus camera, which provides extremely high sensitivity and resolving power due to the large 8.2 meter mirror of the Subaru Telescope. 
\begin{figure}[htb]
    \centering
    \includegraphics[width = 0.47\textwidth]{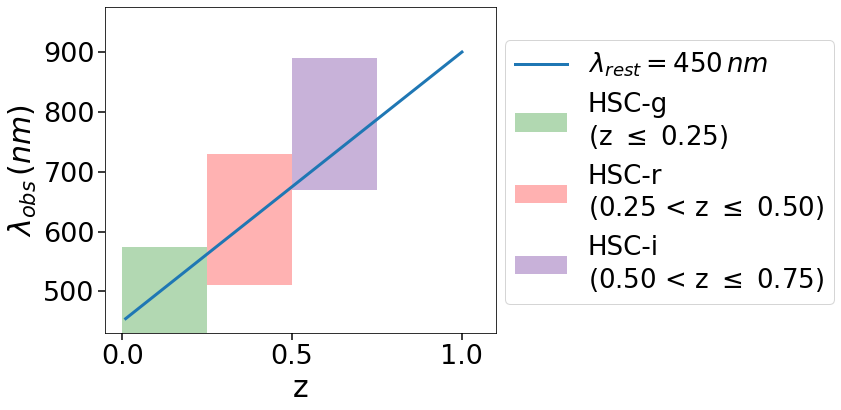}
    \caption{
    The filter used for each redshift bin is shown along with the wavelength range sampled by each filter. The blue line shows where rest-frame $450$ nm emission falls for redshifts labeled on the x-axis. As this figure shows, the chosen filters allow us to consistently perform morphology determination in the rest-frame \gb{}-band.}
    \label{fig:hsc_response}
\end{figure}

In order to have a large but uniform sample of HSC PDR2 galaxies, we focus on the largest volume HSC survey, namely, the Wide layer, which covers $1400$ deg$^2$ to the nominal survey depth in all filters and contains over 450 million primary objects. To consistently perform morphology determination in the rest-frame \gb-band across our entire sample, we use different filters for galaxies at different redshifts, as shown in Figure \ref{fig:hsc_response}. We use \gb-band for $z\leq0.25$, \rb-band for $0.25<z\leq0.50$, and \ib-band for $0.50<z\leq0.75$. Given HSC's typical seeing, all objects with sizes $\lesssim5$\, kpc cannot be resolved beyond $z=0.75$. Therefore, given that the HSC-Wide light profiles of the large majority of galaxies beyond this redshift will be dominated by the PSF, we restrict this work till $z=0.75$ and will explore the application of \gampen{} to higher redshift deeper HSC data in future work. 

In order to select galaxies, we used the PDR2 galaxy catalog produced using forced photometry on coadded images. The HSC data has been reduced using a pipeline built on the prototype pipeline developed by the Rubin Observatory's Large Synoptic Survey Telescope's Data Management system, and we refer the interested reader to \citet{hsc_pipeline} for more details. We use the \texttt{extendedness\_value} flag only to select extended sources. The \texttt{extendedness\_value} flag relies on the difference between the Composite Model (CModel) and PSF magnitudes to separate galaxies from stars, and contamination (from stars) increases sharply for $m > 23$ for median HSC seeing in the Wide layer as outlined \href{https://hsc-release.mtk.nao.ac.jp/doc/index.php/stargalaxy-separation-2/}{here}\footnote{\href{https://hsc-release.mtk.nao.ac.jp/doc/index.php/stargalaxy-separation-2/}{https://hsc-release.mtk.nao.ac.jp/doc/index.php/stargalaxy-separation-2/}}. Thus, for each redshift bin, we  select galaxies with magnitude $m < 23$ (in the appropriate band). The full query to download the data in each redshift bin is available in Appendix \ref{ap:data}.

\begin{deluxetable}{ccccc}[htbp]
\tablecaption{Data Characteristics  \label{tab:data}}
\tablecolumns{5}
\tablehead{
\colhead{Sample} & \colhead{Redshift} & \colhead{Number} & \colhead{Imaging} & \colhead{Spectra} \\
\colhead{Name} & \colhead{} & \colhead{} & \colhead{Band} & \colhead{}}
\startdata
    \hline
    \hline
    Low-z & $z\leq0.25$ & 820,566 & \gb{} & 7.4\% \\
    Mid-z & $0.25 < z\leq0.50$ & 2,937,871 & \rb{} & 2.4\% \\
    High-z & $0.50 < z\leq0.75$ & 4,247,208 & \ib{} & 1.6\% \\
\enddata
\end{deluxetable}

\begin{figure}[htb]
    \centering
    \includegraphics[width = 0.35\textwidth]{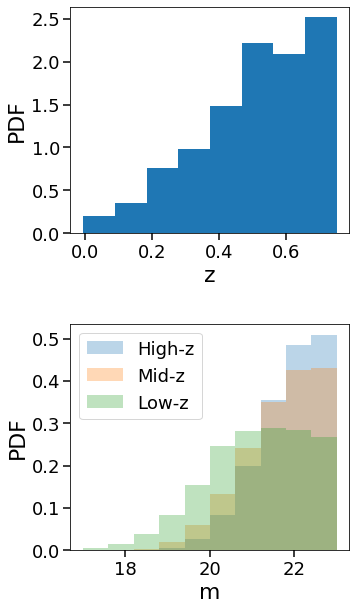}
    \caption{Redshift ({\it top}) and magnitude ({\it bottom}) distributions for the $\sim8$ million galaxies used in this study.
    We used spectroscopic redshifts when available and high-quality photometric redshifts otherwise. The spectroscopic completeness
    of each sub-sample is shown in Table \ref{tab:data}.}
    \label{fig:z_mag_distr}
\end{figure}

We use spectroscopic redshifts when available ($\sim 2.5\%$) and high-quality photometric redshifts otherwise. The spectroscopic redshifts were collated by the HSC Subaru Strategic Program (SSP) team from a wide collection of spectroscopic surveys -- zCOSMOS DR3 \citep{lilly_09}, UDSz \citep{bradshaw_13}, 3D-HST \citep{momcheva_16}, FMOS-COSMOS \citep{silverman_15}, VVDS \citep{lefevre_13}, VIPERS PDR1 \citep{garilli_14}, SDSS DR12 \citep{alam_15}, GAMA DR2 \citep{liske_15}, WiggleZ DR1 \citep{drinkwater_10}, DEEP2 DR4 \citep{newman_13}, PRIMUS DR1 \citep{cool_13}, and VANDELS DR2 \citep{vandels}. The photometric redshifts in HSC PDR2 were calibrated based on the above spectroscopic sample and were calculated using the Direct Empirical Photometric code (DEmP; \citeauthor{demp} \citeyear{demp}), an empirical
quadratic polynomial photometric redshift fitting code, and Mizuki \citep{mizuki}, a Bayesian template fitting code. For an extended description, we refer the interested reader to \citet{photoz_hsc_pdr2}. To remove galaxies with unsecure photometric redshifts, we set the \texttt{photoz\_risk\_best} parameter $<0.1$, which controls the risk of the photometric redshift being outside the range $z_{\mathrm{true}} \pm 0.15(1+z_{\mathrm{true}})$, with 0 being extremely safe to 1 being extremely risky. Using the \texttt{cleanflags\_any} flag, we further excluded objects flagged to have any significant imaging issues (in the relevant band) by the HSC pipeline. This flag can be triggered by a wide range of issues; however, for $\sim 80\%$ of cases, the flag was triggered by cosmic ray hits, as shown in Appendix \ref{ap:data}. We checked and confirmed that none of the above cuts significantly modified the redshift or magnitude distribution of our galaxy sample. 

\begin{figure*}[htb]
    \centering
    \includegraphics[width = 0.9\textwidth]{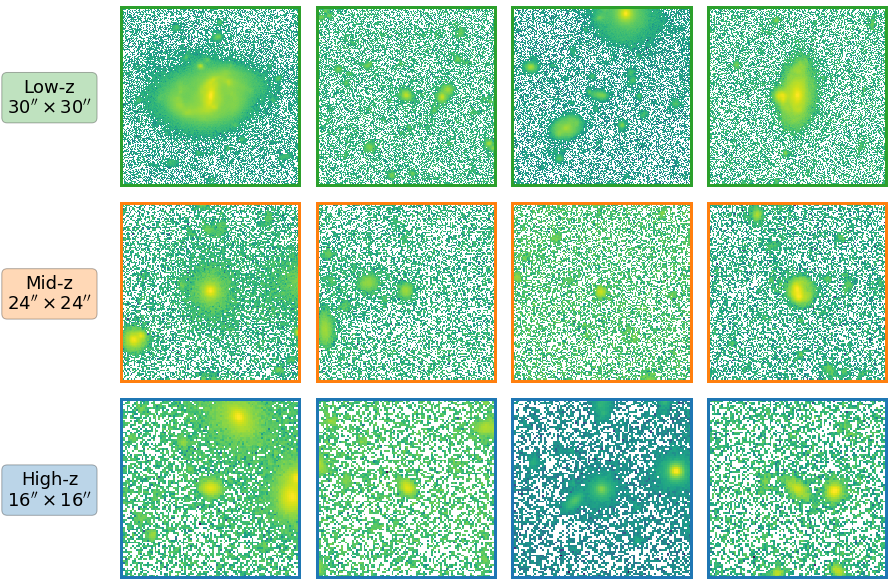}
    \caption{Four randomly chosen galaxy cutouts are shown here for each redshift bin, with the object of interest at the center of each cutout. Note that most of these cutouts have secondary objects in the frame, which can often cause ML algorithms to produce spurious classifications. \gampen{} uses a Spatial Transformer Network to crop most secondary objects out of the frame 
    (see \S 3).}
    \label{fig:real_data_eg}
\end{figure*}

The above process resulted in the selection of $\sim 8$ million galaxies, with $\sim1$ million, $\sim3$ million, $\sim4$ million galaxies in the low-z, mid-z, and high-z bins, as shown in Table \ref{tab:data}. The magnitude and redshift distribution of the data is shown in Figure \ref{fig:z_mag_distr}. Using the \href{https://hsc-release.mtk.nao.ac.jp/das_cutout/pdr2/}{HSC Image Cutout Service}\footnote{\href{https://hsc-release.mtk.nao.ac.jp/das_cutout/pdr2/}{https://hsc-release.mtk.nao.ac.jp/das\_cutout/pdr2/}}, we downloaded cutouts for each galaxy with sizes of $30\arcsec$, $24\arcsec$, $16\arcsec$ for the low-, mid-, and high-z bins, respectively. These sizes are large enough that they should capture all objects in the relevant bin. Using the results of our light-profile fitting analysis on a sub-sample, as outlined in \S \ref{sec:galfitting}, we expect $\sim99.5\%$ of our sample to have a downloaded cutout size $\geq10\times$(size of the major axis of the galaxy). Figure \ref{fig:real_data_eg} shows randomly chosen examples of galaxy cutouts for each redshift bin.

\subsection{Simulated Galaxies for Initial Training} \label{sec:sim_data}

As outlined later in \S \ref{sec:sim_training}, we perform \gampen{}'s initial round of training on mock galaxy image cutouts, simulated to match HSC observations in the appropriate band. To generate mock images, we used GalSim \citep{Rowe2015GalSim:Toolkit}, the modular galaxy image simulation toolkit. GalSim has been extensively tested and shown to yield very accurate rendered images of galaxies. We simulated 150,000 galaxies in each redshift bin, with a mixture of both single and double-component galaxies, in order to have a diverse training sample. To be exact, $75\%$ of the simulated galaxies consisted of both bulge and disk components, while the remaining $25\%$ had either a single disk or a bulge.

\begin{deluxetable*}{c|cccccc}[htbp]
\tablecaption{Parameter Ranges of Simulated Galaxies  \label{tab:sim_para}}
\tablecolumns{7}
\tablehead{
\colhead{Sample} & {Component} & \colhead{\sersic{} Index} & \colhead{Half-Light Radius} & \colhead{Flux} & \colhead{Axis Ratio} & \colhead{Position Angle} \\ 
\colhead{Name} & \colhead{Name} & \colhead{} & \colhead{(arcsec)} & \colhead{(ADUs)} & \colhead{} & \colhead{(degrees)}
}
\startdata
    \hline
    \hline
    \multirow{5}{*}{Low-z} & \multicolumn{6}{c}{Single-Component Galaxies} \\
    & & 0.8 - 1.2 | 3.5 - 5.0\tablenotemark{a} & 0.1 - 5.0 & 30 - $1.35\times10^5$ & 0.25 - 1.0 & $-90.0$ - $90.0$ \\
    & & & & ($m_g \sim $ 14 - 23) & & \\
    \cline{2-7}
    & \multicolumn{6}{c}{Double-Component Galaxies} \\
    & Disk & 0.8 - 1.2 & 0.1 - 5.0 & 0.0 - 1.0\tablenotemark{b} & 0.25 - 1.0 & $-90.0$ - 90.0\\
    & Bulge & 3.5 - 5.0 & 0.1 - 3.0 & 1.0 - Disk. Comp.\tablenotemark{b} & 0.25 - 1.0 &  Disk Comp. $\pm\,[0,15]$\tablenotemark{c} \\
    \hline
    \hline
    \multirow{5}{*}{Mid-z} & \multicolumn{6}{c}{Single-Component Galaxies} \\
    & & 0.8 - 1.2 | 3.5 - 5.0\tablenotemark{a} & 0.1 - 3.0 & 30 - $3\times10^4$ & 0.25 - 1.0 & $-90.0$ - $90.0$ \\
    & & & & ($m_r \sim $ 15.5 - 23) & & \\
    \cline{2-7}
    & \multicolumn{6}{c}{Double-Component Galaxies} \\
    & Disk & 0.8 - 1.2 & 0.1 - 3.0 & 0.0 - 1.0\tablenotemark{b} & 0.25 - 1.0 & $-90.0$ - 90.0\\
    & Bulge & 3.5 - 5.0 & 0.1 - 2.0 & 1.0 - Disk. Comp.\tablenotemark{b} & 0.25 - 1.0 &  Disk Comp. $\pm\,[0,15]$\tablenotemark{c} \\
    \hline
    \hline
    \multirow{5}{*}{High-z} & \multicolumn{6}{c}{Single-Component Galaxies} \\
    & & 0.8 - 1.2 | 3.5 - 5.0\tablenotemark{a} & 0.1 - 2.0 & 30 - $2\times10^4$ & 0.25 - 1.0 & $-90.0$ - $90.0$ \\
    & & & & ($m_i \sim $ 16 - 23) & & \\
    \cline{2-7}
    & \multicolumn{6}{c}{Double-Component Galaxies} \\
    & Disk & 0.8 - 1.2 & 0.1 - 2.0 & 0.0 - 1.0\tablenotemark{b} & 0.25 - 1.0 & $-90.0$ - 90.0\\
    & Bulge & 3.5 - 5.0 & 0.1 - 1.5 & 1.0 - Disk. Comp.\tablenotemark{b} & 0.25 - 1.0 &  Disk Comp. $\pm\,[0,15]$\tablenotemark{c} \\
\enddata
\tablenotetext{a}{The single-component galaxies are equally divided between galaxies with a \sersic{} index between 0.8 - 1.2 and galaxies with a \sersic{} index between 3.5 - 5.0.}
\tablenotetext{b}{Fractional fluxes are noted here. The bulge flux fraction is chosen such that for each simulated galaxy it is added with the disk flux fraction to give $1.0$. The total flux of the galaxies is varied between the values given in the top-row for each sample}
\tablenotetext{c}{The bulge position angle differs from the disk position angle by a randomly chosen value between $-15$ and $+15$ degrees.}
\tablecomments{The above table shows the ranges of the various \sersic{} profile parameters used to simulate mock HSC cutouts. $75\%$ of the simulated galaxies have both disk and bulge components, and the remainder has either a disk or a bulge component. All the simulation parameters are drawn from uniform distributions.}
\end{deluxetable*}

For both the bulge and disk components, we used the \sersic{} profile, and the parameters required to generate the \sersic{} profiles were drawn from uniform distributions over ranges given in Table \ref{tab:sim_para}. For the disk and bulge components, we allow the \sersic{} index to vary between $0.8-1.2$ and $3.5-5.0$, respectively. We chose to have varying \sersic{} indices as opposed to fixed values for each component in order to have a training set with diverse light profiles. Note that the single-component galaxies were included in the simulations to have some examples of galaxies that are purely disk-dominated (i.e., no bulge component) and some that are purely bulge-dominated (i.e., no disk component). Thus, the \sersic{} indices chosen for the single-component galaxies mirror the values chosen for the disk and bulge components in the double-component galaxies.

The parameter ranges for fluxes, and half-light radii are quite expansive and are representative of most galaxies at the appropriate redshift range. To obtain these parameter ranges, we first start with a set of parameters that represent most local galaxies \citep{binney_and_merrifield}. Thereafter, we redshift these parameters for each galaxy using Planck18 cosmology ($H_0=67.7$ km/s/Mpc, \citealp{planck18}) and the appropriate pixel scale.

\begin{figure}[htb]
    \centering
    \includegraphics[width = 0.47\textwidth]{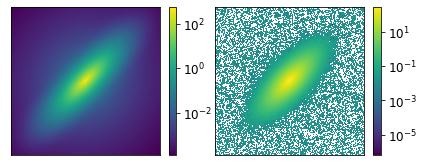}
    \caption{Two stages of simulating an HSC galaxy. (\textit{Left}): A randomly chosen two-dimensional light profile generated by GalSim. (\textit{Right}): The same image after PSF convolution and noise addition. The white pixels represent (small) negative values that arise from the process of noise addition.}
    \label{fig:sim_process}
\end{figure}

To make the two-dimensional light profiles generated by GalSim realistic, we convolved these with representative point-spread functions (PSFs) downloaded from the HSC survey. We also added realistic noise using one-thousand $2\arcsec\times2\arcsec$ ``sky objects" from the HSC PDR2 Wide field. Sky objects are empty regions identified by the HSC pipeline that are outside object footprints and are recommended for being used in blank-sky measurements. For PSF convolution and noise addition, we follow the procedure detailed in \citet{gampen_software_paper} and refer the interested reader to that work for more details. Figure \ref{fig:sim_process} shows a randomly chosen simulated light profile and the corresponding image cutout generated after PSF convolution and noise addition. All the simulated galaxy images in each redshift bin were chosen to have cutout sizes equal to their real data counterparts, as outlined in \S \ref{sec:hsc_data}. 

We would like to note that even after PSF convolution and noise addition, our simulated galaxies are only semi-realistic and do not account for many specific features seen in real data (e.g., spiral arms, knots, non-classical bulges, etc.). The primary goal of the simulation dataset is to provide a large corpus of images on which the initial training can be done. The second step of fine-tuning \gampen{} on real data (described in \S \ref{sec:transfer_learning}) ensures that the framework also learns about the existence of features in the real data that are missed by the simulations.

\section{Brief Introduction to \gampen{}} \label{sec:gampen}
The Galaxy Morphology Posterior Estimation Network \citep[\gampen{};][]{gampen_software_paper} is a novel machine learning framework that can predict posterior distributions for a galaxy's bulge-to-total light ratio ($L_B/L_T$), effective radius ($R_e$), and flux ($F$). In this section, we provide a brief introduction to \gampen{}; however, for a complete understanding of \gampen{}'s architecture and how it predicts posteriors, we refer the reader to Appendix \ref{ap:gampen} and \citet{gampen_software_paper}.

The architecture of \gampen{} consists of an upstream Spatial Transformer Network (STN) module followed by a downstream Convolutional Neural Network (CNN) module. The upstream STN is used to automatically crop each input image frame to an optimal size before morphology determination. Based on the input image, the STN predicts the parameters of an affine transformation which is then applied to the input image. The transformed image is then passed onto the downstream CNN, which estimates the joint posterior distributions of the morphological parameters. Because the transformation we use is differentiable, the STN can be trained using standard backpropagation along with the downstream CNN without any additional supervision.

The inclusion of the STN in the framework greatly reduces the amount of time spent on data pre-processing as we do not have to worry about making cutouts of the ``proper size"---an inherently difficult task since most of the galaxies in our sample have not been morphologically analyzed before. More importantly, the STN greatly reduces the chance of spurious results by cropping most secondary objects out of frame (see \S \ref{sec:stn_performance}).

\begin{figure*}[htb]
    \centering
    \includegraphics[width
    =\textwidth]{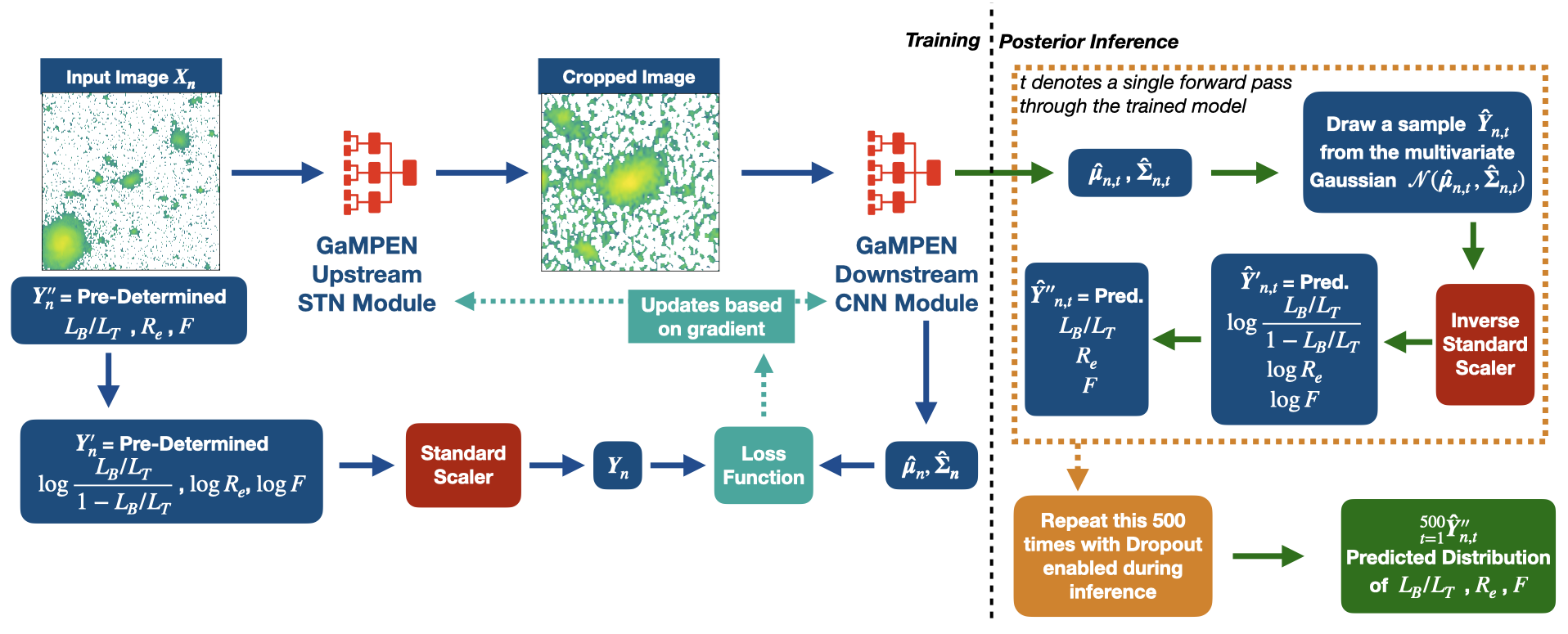}
    \caption{Diagram outlining the training (\textit{left}) and posterior inference (\textit{right}) phases of the \gampen{} workflow. Training consists of feeding galaxies (with pre-determined parameter values) through the STN and CNN modules, minimizing the loss function using Stochastic Gradient Descent. 
    During this process, we re-scale the variables described in the text, and return them to the original variable space during inference.
    After the STN+CNN networks are trained, the posterior inference step consists of 500 forward passes with dropout enabled for each galaxy image. We draw a sample from the predicted multivariate Gaussian distribution during each forward pass, and the collection of these samples gives us the predicted posterior distribution.}
    \label{fig:posterior_pred_workflow}
\end{figure*}

Two primary sources of error contribute to the uncertainties in the parameters predicted by \gampen{}. The first arises from errors inherent to the input imaging data (e.g., noise and PSF blurring), and this is commonly referred to as aleatoric uncertainty. The second source of error comes from the limitations of the model being used for prediction (e.g., the number of free parameters in \gampen{}, the amount of training data, etc.); this is referred to as epistemic uncertainty. 

For every given input image, \gampen{} predicts a multivariate Gaussian distribution $\mathcal{N}(\boldsymbol{\mu}, \boldsymbol{\Sigma})$, with mean $\boldsymbol{\mu}$ and covariance matrix $\boldsymbol{\Sigma}$. Although we would like to use \gampen{} to predict aleatoric uncertainties, the covariance matrix, $\boldsymbol{\Sigma}$, is not known {\it a priori}. Instead, we train \gampen{} to learn these values by minimizing the negative log-likelihood of the output parameters for the training set. The covariance matrix here represents the aleatoric uncertainties in \gampen{}'s predictions.

In order to obtain epistemic uncertainties, we use the Monte-Carlo Dropout technique \citep{Srivastava2014Dropout:Overfitting}, wherein during inference, each image is passed through the trained networks multiple times. During each forward pass, random neurons from the network are removed according to a Bernoulli distribution, i.e., individual nodes are set to zero with a probability, $p$, known as the dropout rate.

The entire procedure used to estimate posteriors is summarized in Figure \ref{fig:posterior_pred_workflow}. Once \gampen{} has been trained, we feed each input image, $\boldsymbol{\hat{X}}_n$, 500 times into the trained model with dropout enabled. During each iteration, we collect the predicted set of $\left(\hat{\boldsymbol{\mu}}_{n,t},\boldsymbol{\hat{\Sigma}}_{n,t}\right)$ for the $t^{th}$ forward pass. Then, for each forward pass, we draw a sample $\boldsymbol{\hat{Y}_{n,t}}$ from the multivariate normal distribution $\mathcal{N}\left(\boldsymbol{\hat{\mu}}_{n,t},\boldsymbol{\hat{\Sigma}}_{n,t}\right)$. The distribution generated by the collection of all 500 forward passes, $\boldsymbol{\hat{Y}_{n}}$,
represents the predicted posterior distribution for the test image $\boldsymbol{\hat{X}}_n$. The different forward passes capture the epistemic uncertainties, and each prediction in this sample also has its associated aleatoric uncertainty represented by $\boldsymbol{\hat{\Sigma}_{n,t}}$. Thus the above procedure allows us to incorporate both aleatoric and epistemic uncertainties in \gampen{}'s predictions.

\section{Training \gampen{}} \label{sec:method}
 
Since most of the galaxies described in \S \ref{sec:hsc_data} have not been morphologically analyzed before, we devise a method to train \gampen{} with only $0.5\%$ of our sample of real galaxies. In order to achieve this, we first train \gampen{} using the simulated galaxies described in \S \ref{sec:sim_data}. Thereafter, we apply ``transfer learning", wherein we fine-tune the already trained network using a small sample of real galaxies, analyzed using GALFIT \citep{galfit}. 

\begin{figure*}[htb]
    \centering
    \includegraphics[width
    =0.7\textwidth]{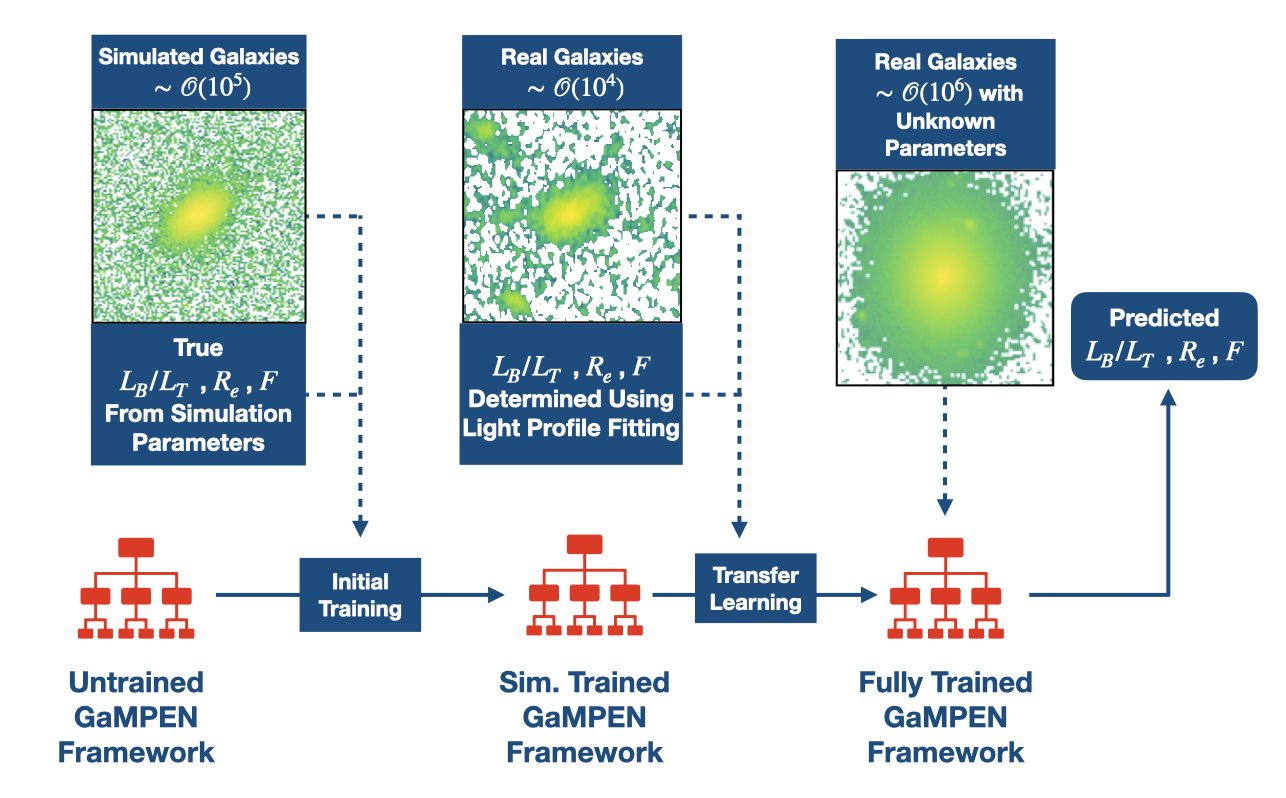}
    \caption{Diagram outlining the different stages of training \gampen{}. We first train \gampen{} using simulated light profiles described in \S \ref{sec:sim_data}. Thereafter, we fine-tune the simulation-trained framework using $0.5\%$ of our real data sample, for which we pre-determined the morphological parameters using light-profile fitting, as described in \S \ref{sec:galfitting}. Finally, we process all the $\sim 8$ million galaxies in our dataset through the trained \gampen{} framework to obtain estimates of their morphological parameters and associated uncertainties.}
    \label{fig:full_gampen_workflow}
\end{figure*}

The process of training \gampen{} consists of the following steps, summarized in Figure \ref{fig:full_gampen_workflow}.
\begin{itemize}
\item Simulating galaxies corresponding to the target data set (described previously in \S \ref{sec:sim_data}).
\item Initial training of \gampen{} on the above simulated images (described further in \S \ref{sec:sim_training}).
\item Fine-tuning \gampen{} using a small fraction of the real dataset. For this, we used $\sim0.5\%$ of the HSC data described in \S \ref{sec:hsc_data}. This process is known as transfer learning and is described further in \S \ref{sec:transfer_learning}.
\item Testing the performance of the fine-tuned network on a test set of real galaxies. For this, we used $\sim0.25\%$ of the HSC data described in \S \ref{sec:hsc_data}.
\item Processing the remainder of the real data ( $\sim99\%$ of the HSC data described in \S \ref{sec:hsc_data}) through the trained framework.
\end{itemize}

We describe each of the above steps in detail below. 

\subsection{Data Transformations} \label{sec:transformations}

To make the training process more robust against numerical instabilities, we transform the input images and target variables following the steps outlined in \citet{gampen_software_paper}.

Since reducing the dynamic range of pixel values has been found to be helpful in neural network convergence \citep[e.g.,][]{zanisi_21,walmsley_decals,tanaka_22}, we pass all images in our dataset through the arsinh function. For all the target variables, we first apply the logit transformation to $L_B/L_T$ and log transformations to $R_e$ and $F$: 

\begin{equation}
\boldsymbol{Y_n'} = f''(\boldsymbol{Y}_n'') = \left( \log \frac{L_B/L_T}{1 - L_B/L_T}, \log R_e, \log F \right) ,
\label{eq:transformation_f''}
\end{equation} 

\noindent
where $\boldsymbol{Y}_n'' = [{L_B/L_T,R_e,F}]$ is the target set of variables before transformation and $f''$ is how we will refer to the transformation in Equation\,\ref{eq:transformation_f''}. Next, we apply the standard scalar transformation to each parameter (calibrated on the training data), which amounts to subtracting the mean value of each parameter and scaling its variance to unity. These two transformations ensure that all three variables have similar numerical ranges and prevent variables with larger numerical values from making a disproportionate contribution to the loss function.

Post training, during inference, we apply the inverse of the standard scalar function (with no re-tuning of the mean or variance), followed by the inverse of the logit and log transformations, $f''^{-1}$, as indicated in Figure\,\ref{fig:posterior_pred_workflow}. Besides transforming the variables back to their original space, these final transformations also ensure that the predicted values always conform to physically meaningful ranges ($0 \leq L_B/L_T \leq 1$; $R_e > 0$; $F > 0$).

\subsection{Initial Training of \gampen{} on simulated galaxies} \label{sec:sim_training}
The purpose of training \gampen{} initially on simulations is two-fold. Firstly, it greatly reduces the amount of real data needed for training the framework. Secondly, since simulated galaxies are the only situation where we have access to the ``ground-truth" morphological parameters, they provide the perfect dataset to assess \gampen{}'s typical accuracy for the different output parameters. 

In \citet{gampen_software_paper}, we extensively tested and reported on \gampen{}'s performance on simulated HSC $z \leq 0.25$ \gb{}-band galaxies. Here, we extend that to include simulated \rb{}-band $0.25 < z \leq 0.5$ and \ib{}-band $0.50 < z \leq 0.75$  galaxies. Out of the 150,000 galaxies simulated in each z-bin, we use $\mathbf{70\%}$ to train the framework and $15\%$ as a validation set to determine the optimum value for various hyper-parameters (such as learning rate, batch size, etc.). Thereafter, we use the remaining $15\%$, which the framework has never encountered before, to evaluate the performance of the trained framework. 

We train \gampen{} by minimizing its loss function using Stochastic Gradient Descent. The different hyper-parameters that need to be tuned are: the learning rate (the step size during gradient descent), momentum (acceleration factor used for faster convergence), the strength of L2 regularization (the degree to which larger weight values are penalized), and batch size (the number of images processed before weights and biases is updated). To choose these hyper-parameters, we trained \gampen{} with different sets of hyper-parameters and chose the ones that resulted in the lowest value for the loss function on the validation set. The final 
hyper-parameters for the trained models are given in Appendix \ref{ap:sec:trained_models}.

\begin{figure*}[htb]
    \centering
    \includegraphics[width = 0.9\textwidth]{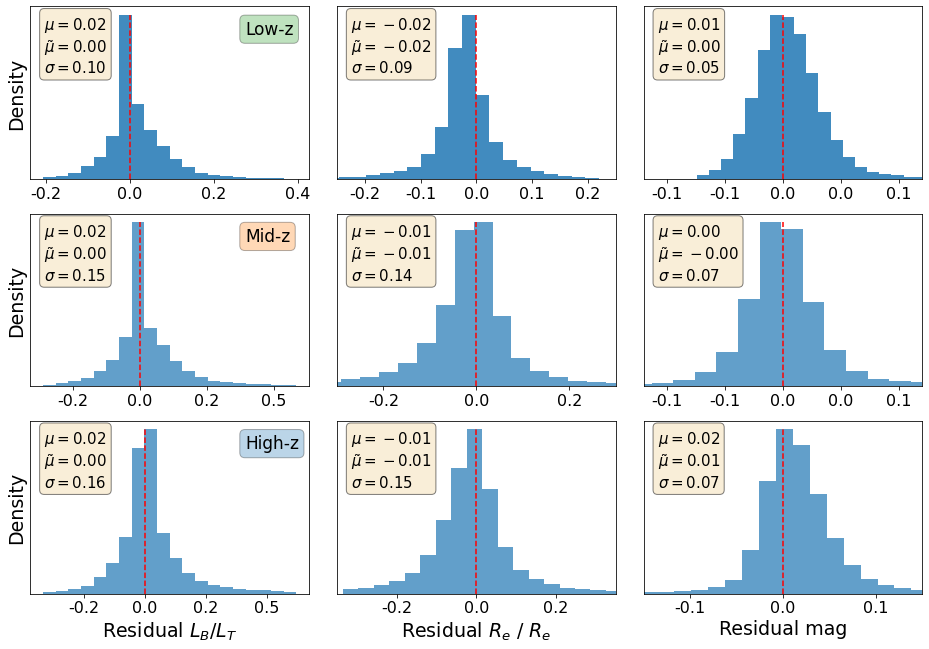}
    \caption{Histograms of residuals for simulated galaxies (in the test set) across all three redshift bins. We define the residuals as the difference between the true value and the most probable value predicted by \gampen{}. The dashed vertical line represents $x = 0$, denoting cases with perfectly recovered parameter values. 
    The mean ($\mu$), median ($\tilde{\mu}$), and standard deviation ($\sigma$) of each residual distribution are listed in each panel.}
    \label{fig:resi_sim_all_z}
\end{figure*}

In order to test the robustness of our simulation-trained framework, we compare the predictions made by \gampen{} on the test set to the true values determined from the simulation parameters. The results are similar across all redshift bins and closely follow what we determined previously in \citet{gampen_software_paper}. The histograms of residuals for \gampen{}'s output parameters are shown in Figure \ref{fig:resi_sim_all_z} across all three redshift bins. Note that to make all three parameters dimensionless, we report the (Residual $R_e$)/$R_e$, instead of simply Residual $R_e$. For each histogram, we also show the mean ($\mu$), median ($\tilde{\mu}$), and standard deviation $(\sigma)$. The mean and median help demonstrate that the distributions are all centered around zero, and the standard deviation indicates the value of the ``typical error" made by the framework (i.e., $68.27\%$ of the time, \gampen{}'s prediction errors are less than this value).

Note that for the simulated data, the typical error made by \gampen{} increases with redshift. This is expected given that in our simulations, galaxies in the higher redshift bins are preferentially smaller, fainter, and have lower signal-to-noise ratios than their lower redshift counterparts---thus, these galaxies are harder to analyze morphologically for any image processing algorithm. However, for all the parameters across all redshift bins, the \gampen{} error is typically always $\lesssim 15\%$ for the simulated sample.

We would like to note that \gampen{} does not explicitly predict the number of components a galaxy has, but we performed an analysis of its relative performance on single- and double-component galaxies in \citet{gampen_software_paper} (see Figure 15 therein). We found that accurately determining $L_B/L_T$ is more challenging for double-component galaxies (compared to single-component galaxies); and becomes even more difficult when one of the components strongly dominates the other.

\subsection{Transfer Learning using Real Data} 
\label{sec:transfer_learning}

Transfer Learning as a data-science concept has been around since the late 1990s \citep[e.g.,][]{blum_1998} and involves taking a network trained on a particular dataset and optimized for a specific task, and re-tuning/fine-tuning the weights and biases for a slightly different task or dataset.  In \citet{gamornet_paper}, we introduced the concept of training on simulated light-profiles and then transfer learning using real data for galaxy morphology analysis. Here, we employ the same idea, taking the networks trained on the simulated galaxies and then employing transfer learning using $\sim15,000$ real HSC-Wide galaxies in each redshift bin.

\begin{figure*}[htb]
    \centering
    \includegraphics[width = 0.9\textwidth]{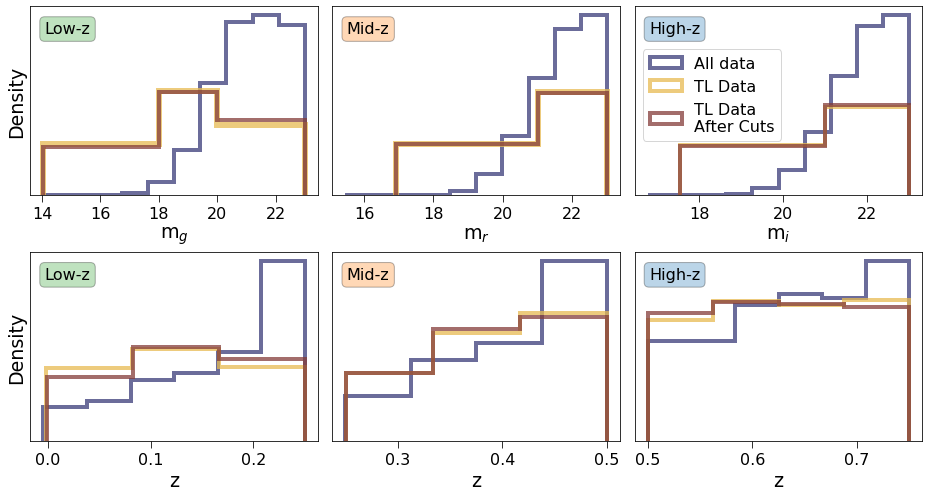}
    \caption{Magnitude and redshift distributions for all galaxies in each redshift bin, plotted along with the galaxies selected for transfer learning (before and after applying quality cuts, as described in \S \ref{sec:galfitting}). Note that we plot density on the y-axis, not the number of samples. The total number of galaxies used for transfer learning is $\sim0.5\%$ of all the galaxies in our dataset. The relative density of some magnitude bins is higher than others (e.g., $18 < m \leq 20$ for low-z) because they span a smaller range while having roughly the same number of galaxies as the other bins.}
    \label{fig:tl_para_dist}
\end{figure*}

In order to select a sample of galaxies to use for transfer learning, we start with the galaxies summarized in Table \ref{tab:data}. Note that what matters in training a CNN is not matching the observed distributions of the simulation parameters; rather, it is spanning the full range of those parameters with high fidelity. Having too many of a particular type---even if that is the reality in real data---can result in lower accuracy for minority populations (e.g., \citealp{gamornet_paper}). As seen in Figure \ref{fig:z_mag_distr}, the sample for all three redshift bins is heavily biased towards fainter galaxies. To ensure that \gampen{} gets to train on enough bright galaxies, we split the mid- and high-z samples into two sub-samples: $m \leq 21$ and $m > 21$. For the low-z sample, since it has a more substantial tail towards the lower m values (compared to the mid- and high-z samples), we use  three sub-samples---$m \leq 18$, $18 < m \leq 20$, $m > 20$. We select 20,000 galaxies from each redshift bin, making sure to sample equally across the magnitude bins mentioned above. Thereafter, we determine their morphological parameters using light-profile fitting, as described later in \S \ref{sec:galfitting}. The magnitude and redshift distributions of the selected galaxies are shown in Figure \ref{fig:tl_para_dist}. As seen from the figure, the transfer learning sample has sufficiently large numbers of examples from all parts of the parameter space. This empowers us to optimize \gampen{} for the full range of galaxy morphologies.

Out of the 20,000 galaxies selected in each redshift bin, we use $75\%$ for fine-tuning the simulation trained \gampen{} models and another $5\%$ for selecting the various hyper-parameters to be used during the transfer learning process. The remaining $20\%$ is used to evaluate the performance of the fully trained \gampen{} frameworks in \S \ref{sec:results}.

In order to artificially augment the number of galaxies being used for transfer learning, we apply random rotations and horizontal/vertical flips on the galaxies earmarked for training. This takes the effective number of samples used for fine-tuning in each bin from $\sim15,000$ to $\sim90,000$. Using the values determined by light-profile fitting as the labels, we fine-tune the three \gampen{} models trained on simulations. We choose the values of the different hyper-parameters based on the loss computed on the validation set, and the final chosen hyper-parameters for transfer learning are reported in Appendix \ref{ap:sec:trained_models}. 
    
\subsubsection{Fine-Tuning the Dropout Rate} \label{sec:fine_tuning_dropout}

\begin{figure*}[htb]
    \centering
    \includegraphics[width = 0.7\textwidth]{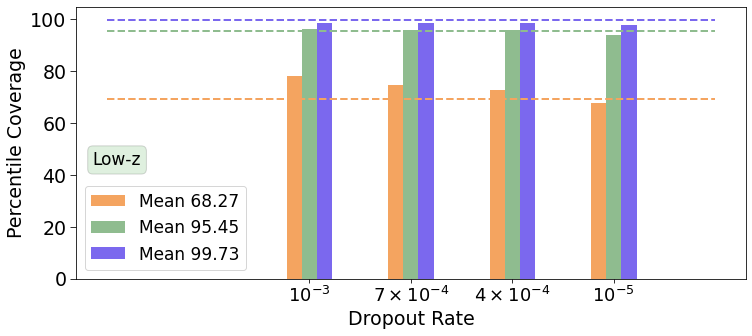}
    \caption{The calculated percentile coverage probabilities for different dropout rates for the low-z bin. Note that the coverage probabilities have been averaged over the three output variables. The coverage probabilities are defined as the percentage of the total test examples where the value determined using light profile fitting lies  within  a  particular confidence interval of the predicted distribution. A dropout rate of $4\times10^{-4}$ leads to coverage probabilities very close to their corresponding confidence levels. A similar process of tuning in the mid-z and high-z bins leads to an optimal dropout rate of $2\times10^{-4}$ in both of them.}
    \label{fig:dropout_calibration}
\end{figure*}

Aside from the hyper-parameters mentioned in \S \ref{sec:sim_training}, there is one more adjustable parameter in \gampen{}---the dropout rate, which directly affects the calculation of the epistemic uncertainties outlined in \S \ref{sec:gampen}. On average, higher dropout rates lead networks to estimate higher epistemic uncertainties. To determine the optimal value for the dropout rate, during transfer learning, we trained variants of \gampen{} with dropout rates from $10^{-3}$ to $10^{-5}$, all with the same optimized values of momentum, learning rate, and batch size mentioned in Appendix \ref{ap:sec:trained_models}.

To compare these models, we calculate the percentile coverage probabilities associated with each model, defined as the percentage of the total test examples where the parameter value determined using light profile fitting lies within a particular confidence interval of the predicted distribution. We calculate the coverage probabilities associated with the $68.27\%$, $95.45\%$, and $99.73\%$ central percentile confidence levels, corresponding to the $1\sigma$, $2\sigma$, and $3\sigma$ confidence levels for a normal distribution. For each distribution predicted by \gampen{}, we define the $68.27\%$ confidence interval as the region on the x-axis of the distribution that contains $68.27\%$ of the most probable values of the integrated probability distribution. To estimate the probability distribution function from the \gampen{} predictions (which are discrete), we use kernel density estimation, which is a non-parametric technique to estimate the probability density function of a random variable. 

We calculate the $95.45\%$ and $99.73\%$ confidence intervals of the predicted distributions in the same fashion. Finally, we calculate the percentage of examples for which the GALFIT-ed parameter values lie within each of these confidence intervals. An accurate and unbiased estimator should produce coverage probabilities equal to the confidence interval for which it was calculated (e.g., the coverage probability corresponding to the $68.27\%$ confidence interval should be $68.27\%$). For every redshift bin, we choose the dropout rate for which the calculated coverage probabilities are the closest to their corresponding confidence levels. This leads to a dropout rate of $4\times10^{-4}$ for the low-z bin, and $2\times10^{-4}$ for the mid- and high-z bins. 

As an example, we show in Figure \ref{fig:dropout_calibration} the coverage probabilities (averaged across the three output variables) for different dropout rates for the low-z sample. As can be seen, higher values of the dropout rate lead to \gampen{} over-predicting the epistemic uncertainties, resulting in too high coverage probabilities. In contrast, extremely low values lead to \gampen{} under-predicting the epistemic uncertainties. For a dropout rate of $4\times10^{-4}$, the calculated coverage probabilities are very close to their corresponding confidence levels, resulting in accurately calibrated posteriors.

\section{Galfitting Galaxies for Transfer Learning \& Validation} \label{sec:galfitting}

\begin{figure*}[htb]
    \centering
    \includegraphics[width = 0.85\textwidth]{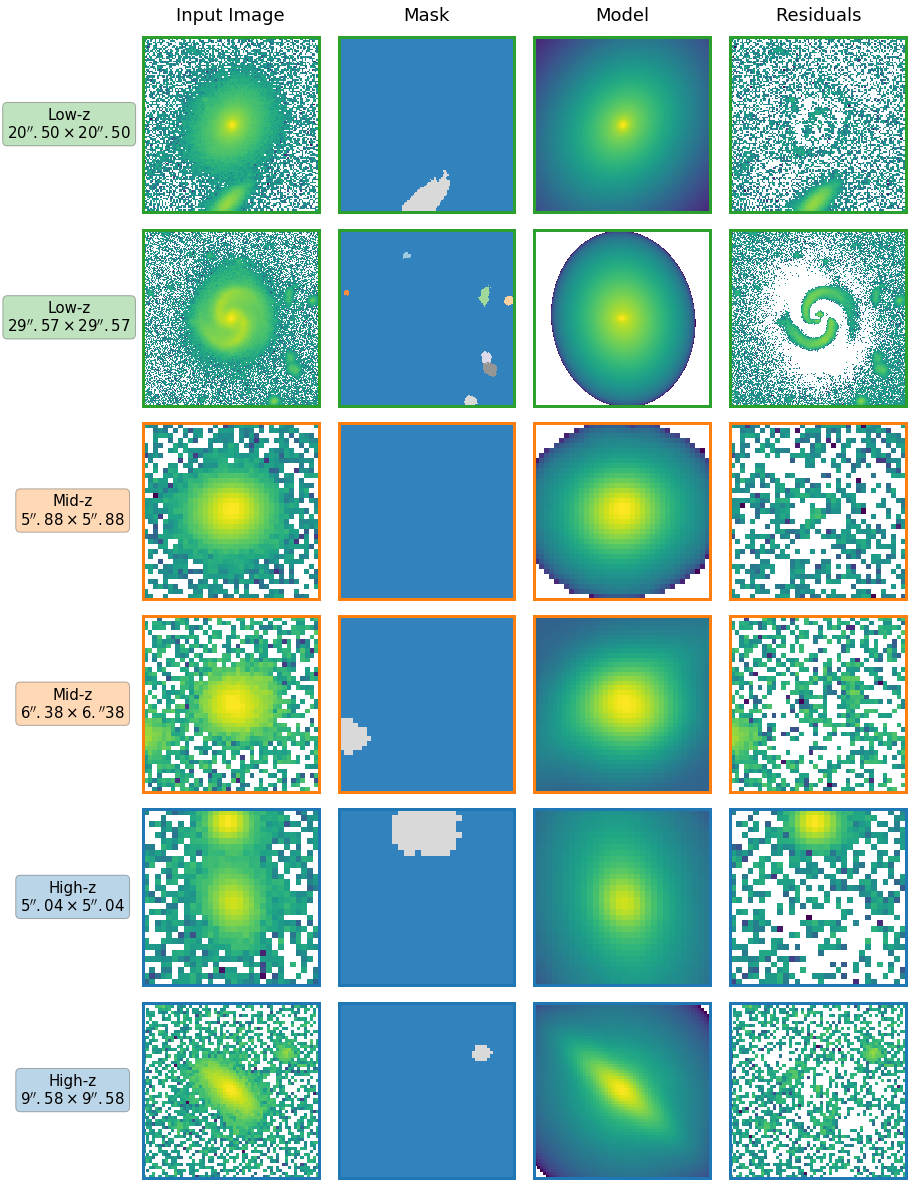}
    \caption{
     Steps used in our light-profile fitting pipeline to determine morphological parameters, for two representative galaxies in each redshift bin. From left to right we show the input image, the  mask generated by Source Extractor, the model generated by GALFIT, and the residuals. Note that since we do not explicitly model any Fourier bending modes or coordinate rotations, we expect features like spiral arms to show up in the residuals, as depicted in the second row.
    }
    \label{fig:galfit_example_fits}
\end{figure*}

\begin{figure*}[htb]
    \centering
    \includegraphics[width = 0.95\textwidth]{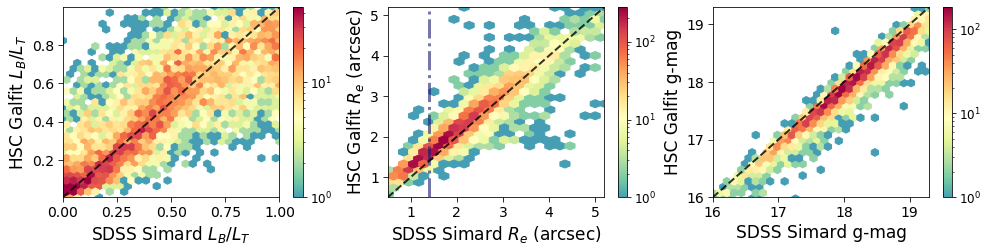}
    \caption{Morphological parameters determined for HSC-imaged galaxies using our light-profile fitting pipeline, versus morphological parameters for the same galaxies determined by \citet{simard_11} based on SDSS imaging. The black dashed diagonal corresponds to perfect agreement. The vertical line in the middle panel shows the median SDSS g-band seeing.}
    \label{fig:gfit_vs_simard}
\end{figure*}

In this section, we describe a semi-automated pipeline that we developed and used to determine the morphological parameters of $\sim60,000$ galaxies ($\sim20,000$ in each z-bin), which are used for transfer learning and to test the efficacy of the trained \gampen{} frameworks.

In order to estimate the parameters, we use GALFIT, which is  a two-dimensional fitting algorithm designed to extract structural components from galaxy images. However, before running GALFIT, we run Source Extractor \citep{s_extract} on all the cutout frames in order to obtain segmentation maps for each input image. We use these segmentation maps to mask all secondary objects present in the cutout frame during light-profile fitting. We also use the Source Extractor estimates of various morphological parameters as the initial starting guesses during light profile fitting. Lastly, we use the Source Extractor estimates to pick a cutout size for each galaxy, which we set to be ten times the effective radius estimated by Source Extractor. 

Using GALFIT, we fit each galaxy with two \sersic{} components:- a disk-component with \sersic{} index, $n=1$, and a bulge component with $3.5 < n < 5.0$. For each galaxy, we perform two consecutive rounds of fitting:- first with some constraints placed on the values of the different parameters and second round with almost no constraints. This is to help the parameters converge quickly during the initial round while still allowing the full exploration of the parameter-space in the subsequent round. In the initial round, we constrain the radius of each galaxy to be between 0.5 - 90 pixels (0.85 - 15.12 arcsec) and the magnitude to be between $\pm7.5$ of the initial value guessed by Source Extractor. We also constrain the difference between the magnitudes of the two components to be between $-7.5$ and $+7.5$ ($0.001 < L_B/L_T < 0.999$), and the relative separation between the centers of the two light profiles to be not more than 30 pixels ($5.04$ arcsec). These constraints are much more expansive than what we expect galaxy parameters to be in this z-range and are similar to what has been used for many previous studies \citep[e.g.,][]{vdw_12,Tuccillo2018DeepFitting}. During the second round of fitting, we only use the constraints on the \sersic{} indices and the relative separation between the two components. After fitting the two components, we calculate the $R_e$ of each galaxy as the radius that contains $50\%$ of the total light in the combined disk + bulge model. Figure \ref{fig:galfit_example_fits} shows the image cutouts, masks, fitted models, and residuals for two typical galaxies chosen from each redshift bin.

After the two rounds of fitting, we excluded galaxies for which GALFIT failed to converge ($\sim3.2\%$, $\sim3.8\%$, and $\sim2.9\%$ for the low-, mid-, and high-z bins, respectively). Thereafter, we visually inspected the fits of a randomly selected sub-sample of $\sim300$ galaxies from each redshift bin. The process of visually inspecting $\sim900$ galaxies helped us to identify three failure modes of our fitting pipeline: i) for some galaxies, GALFIT assigned an extremely small axis ratio to one of the components leading to an unphysical lopsided bulge/disk; ii) for some galaxies, the centroids of the two fitted components were too far away from each other; iii) some galaxies had residuals that were too large. Examples of each of these failure modes are shown in Appendix \ref{ap:sec:galfit_failures}.

\begin{deluxetable}{c|c|c}[htbp]
\tablecaption{GALFIT Quality Cuts \label{tab:galfit_quality_cuts}}
\tablecolumns{3}
\tablehead{
\colhead{Sample} & \colhead{Criteria} & \colhead{Excluded \%} }
\startdata
    \hline
    \hline
    \multirow{3}{*}{Low-z} & (\texttt{problematic\_value\_flags == True})  & \multirow{3}{*}{26.04\%} \\
    & OR (Sep. Dist. $> 2''$) & \\
    & OR (\texttt{max\_iters\_flag == True} & \\ 
    & AND Red. $\chi^2 > 2.5$) & \\
    \hline
    \multirow{3}{*}{Mid-z} & (\texttt{problematic\_value\_flags == True})  & \multirow{3}{*}{32.10\%} \\
    & OR (Sep. Dist. $> 1''$) & \\
    & OR (\texttt{max\_iters\_flag == True} & \\ 
    & AND Red. $\chi^2 > 1.25$) & \\
    \hline
    \multirow{3}{*}{High-z} & (\texttt{problematic\_value\_flags == True})  & \multirow{3}{*}{34.13\%} \\
    & OR (Sep. Dist. $> 0''.7$) & \\
    & OR (\texttt{max\_iters\_flag == True} & \\ 
    & AND Red. $\chi^2 > 1.25$) & \\
\enddata
\end{deluxetable}

In order to get rid of these problematic fits from our GALFIT-ed sample, we use a mixture of GALFIT flags and calculated parameters. Specifically, as summarized in Table \ref{tab:galfit_quality_cuts} and detailed below:

\begin{enumerate}[label=\alph*)]
    \item \texttt{problematic\_value\_flags}\footnote{Refer to Bullet 8 at \href{https://users.obs.carnegiescience.edu/peng/work/galfit/TOP10.html}{\url{https://users.obs.carnegiescience.edu/peng/work/galfit/TOP10.html}}} flags galaxies that have an extremely small axis-ratio ($<0.1$) or radius ($<0.5$ pixels), or any other parameters that caused issues with numerical convergence. 
    \item \texttt{max\_iters\_flag} flags galaxies for which GALFIT quit after reaching the maximum number of iterations (100).
    \item The reduced $\chi^2$ of the fit is poor.
    \item The distance between the centers of the two fitted components is too large.
\end{enumerate}

Table \ref{tab:galfit_quality_cuts} shows the criteria used to exclude fits in each redshift bin. The variation in the thresholds with redshift is to account for the fact that galaxies at higher redshift are preferentially smaller, fainter, and have lower signal-to-noise ratios. Figure \ref{fig:tl_para_dist} shows that the exclusion criteria do not selectively exclude more galaxies from certain regions of the magnitude/redshift parameter space compared to others. In order to determine appropriate thresholds for the various flags above, we balanced excluding too many galaxies against ensuring only good fits are included in the final transfer learning dataset. The choice of thresholds is arbitrary to a certain extent. In order to empower users to retrain \gampen{} using different criteria for their own scientific analysis, we are making public the entire catalog of GALFIT-ed values as outlined in Appendix \ref{sec:ap:data_access}.


\textcolor{black}{To the best of our knowledge, there is no published large catalog of bulge+disk decomposition of HSC galaxies against which we can compare the results of our light-profile fitting pipeline. However, \citet{simard_11} performed bulge+disk decomposition using \gb{} and \rb{}-band imaging from the Sloan Digital Sky Survey (SDSS). We should note that HSC-Wide differs from SDSS in a multitude of ways, with the most significant differences being in median seeing [HSC-Wide: $0.\arcsec79$ compared to SDSS: $1.\arcsec4$ in the g-band] and pixel scale [HSC: 0.168 arcsecs/pixel compared to SDSS: 0.396 arcsecs/pixel]. Thus, we do not expect our analysis to yield the exact same results as that of \citet{simard_11}. However, given that a significant portion of our low-z sample overlaps with that of \citet{simard_11}, it is still useful to compare our results to that of \citet{simard_11}, as the overall trends should agree.}

Using an angular cross-match diameter of $0.15$ arcsec, we cross-matched our low-z GALFIT sample with that of \citet{simard_11} to obtain a sample of $\sim6500$ galaxies. Note that although our HSC sample extends to $g < 23$, the cross-matched galaxies are mostly $g < 20$ due to the shallower depth of the SDSS data. In Figure \ref{fig:gfit_vs_simard}, we compare the results of our light-profile fitting results with that of \citet{simard_11}. The figure shows galaxies in hexagonal bins of roughly equal size, with the number of galaxies in each bin represented according to the colorbar on the right. Note that we are using a logarithmic colorbar to explore the full distribution of galaxies, down to 1 galaxy/bin. \textcolor{black}{Although there are outliers present for all three parameters, and the scatter of the relationship depends on the parameter, our GALFIT-derived parameters strongly correlate with that of \citet{simard_11}. According to the Spearman's rank correlation test \citep[see][for more details]{spearman}, there is a positive correlation for all three variables, and the null hypothesis of non-correlation can be rejected at extremely high significance ($p < 10^{-200}$). The correlation coefficients obtained for $L_B/L_T$, $R_e$, and $F$ are 0.85, 0.95, and 0.98, respectively.} For both our GALFIT predictions and that of \citet{simard_11}, we define $R_e$ for double-component fits as the radius that encompasses $50\%$ of the light from both components combined.

\textcolor{black}{Note that the higher scatter in the $L_B/L_T$ relation is expected, given that bulge+disk decomposition involves constraining two different imaging components simultaneously and is thus more sensitive to algorithmic differences and the differences in imaging quality between SDSS and HSC mentioned above. These differences in imaging quality, data-reduction pipeline, and filters \citep{hsc_filters} might also explain the slight offsets in the $R_e$ and flux measurements. It is also important to note that most of the differences in $R_e$ measurements are at effective radii values smaller than the median SDSS g-band seeing, as seen in the middle panel of Figure \ref{fig:gfit_vs_simard}. } 

\section{Evaluating \gampen{}'s performance} \label{sec:results}
After both rounds of training are complete, we apply the final trained \gampen{} models to all the $\sim 8$ million galaxies outlined in \S \ref{sec:hsc_data}. In this section, we evaluate \gampen{}'s performance to assess the reliability of its predictions.

\subsection{Inspecting the Predicted Posteriors}

\begin{figure*}[htb]
    \centering
    \includegraphics[width = 0.9\textwidth]{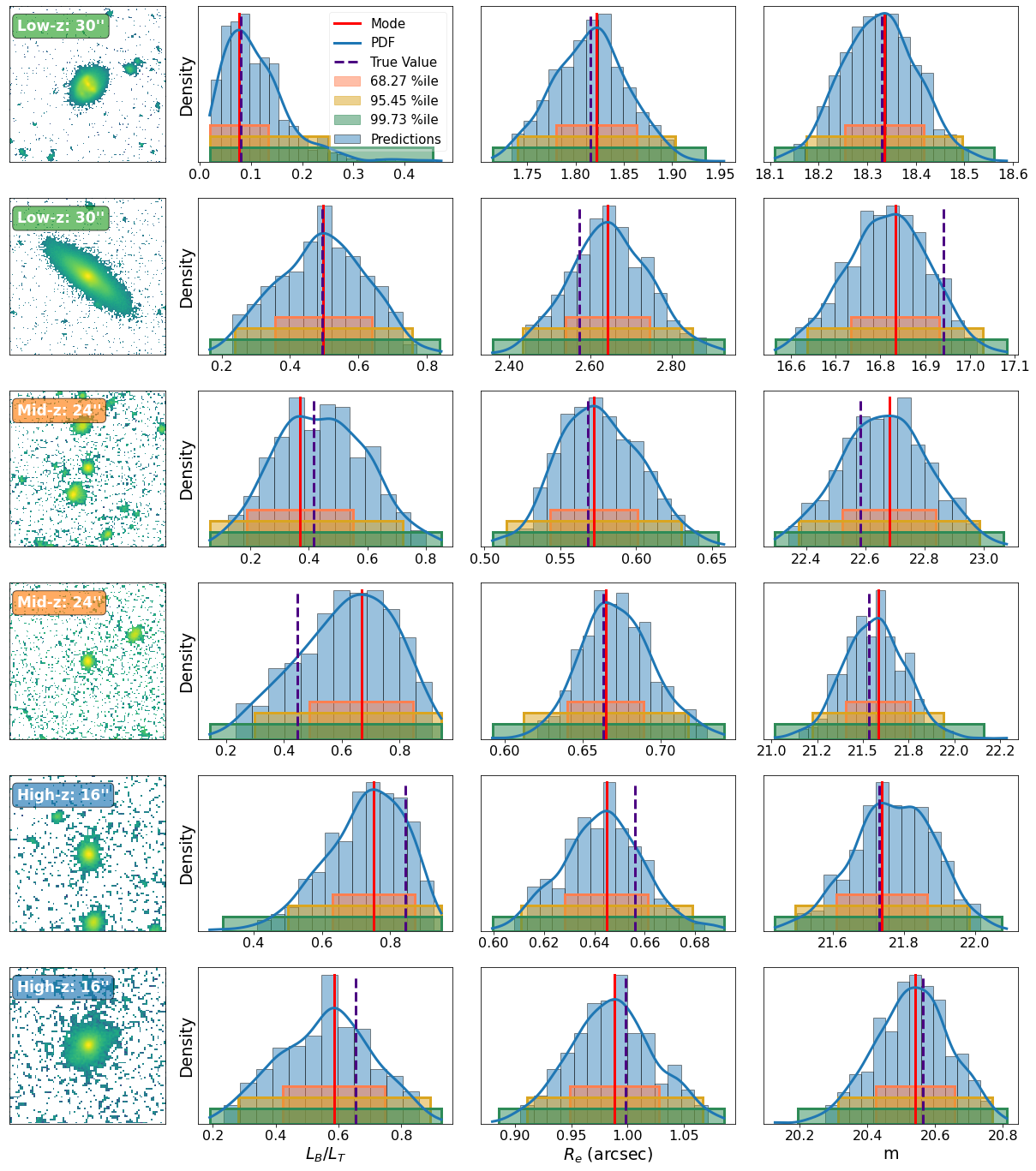}
    \caption{Examples of predicted posterior distributions for two randomly chosen galaxies from each redshift bin. The blue shaded histograms show the predictions from \gampen{}, and the solid blue lines show the associated probability distribution functions estimated by kernel density estimation. These are used to calculate the confidence intervals shown in the figure with pink, yellow, and green shading. The mode (solid red line) shows the most probable value of each morphological parameter. As expected, in most cases, the GALFIT-ed value (dashed black line) lies within the $68.27\%$ confidence interval.}
    \label{fig:example_pred_dists}
\end{figure*}

Using the procedure outlined in \S \ref{sec:gampen} and Figure \ref{fig:posterior_pred_workflow}, we use the trained \gampen{} frameworks to obtain joint probability distributions of all the output parameters for each of the $\sim 8$ million galaxies. Figure \ref{fig:example_pred_dists} shows the marginalized posterior distributions for six randomly selected galaxies; along with the input cutouts fed to \gampen{}. All the predicted distributions are unimodal, smooth, and resemble Gaussian/skewed-Gaussian distributions. For each predicted distribution, the figure also shows the parameter space regions that contain $68.27\%$, $95.45\%$, and $99.73\%$ of the most probable values of the integrated probability distribution. We use kernel density estimation to estimate the probability distribution function (PDF; shown by a blue line in the figure) from the predicted values. The mode of this PDF is what we refer to as the ``predicted value" henceforth. The figure also demonstrates how \gampen{} predicts distributions of different widths based on the input frame (e.g., the $L_B/L_T$ distribution in the second row is wider compared to the first row), and we will explore this in more detail in \S \ref{sec:uncertainties}. 

By design, \gampen{} predicts only physically possible values. This is especially apparent in the $L_B/L_T$ column of rows 1 and 4 of Figure \ref{fig:example_pred_dists}. Note that to achieve this, we do not artificially truncate these distributions. Instead, we use data transformations, as outlined in \S \ref{sec:transformations}. This ensures that the predicted $L_B/L_T$ values are always between 0 and 1. Similarly, we also ensure  that the $R_e$ and $F$ values predicted by \gampen{} are positive through appropriate transformations. 

While performing quality checks on the predicted posteriors of all the $\sim 8$ million galaxies, we noticed that sometimes \gampen{} predicts $R_e$ and F values outside the parameter range on which it was trained (e.g., $m > 23$). It has been shown that while machine learning frameworks are excellent at interpolation, one should be extremely cautious while trying to extrapolate too much beyond the training set \citep[e.g.,][]{candela_09, recht_19, taori_20}. Thus, for each redshift bin, we exclude galaxies with predicted $R_e$ and $F$ values which are outside the upper and lower bounds of the training set by more than 0.5 arcsecs or 0.5 mags, respectively. This led to $\sim6\%$, $\sim2.5\%$, and $\sim1.2\%$ of the data being excluded in the low-, mid-, and high-z bins.

\subsection{Evaluating the STN performance} \label{sec:stn_performance}

\begin{figure*}[htb]
    \centering
    \includegraphics[width = 0.9\textwidth]{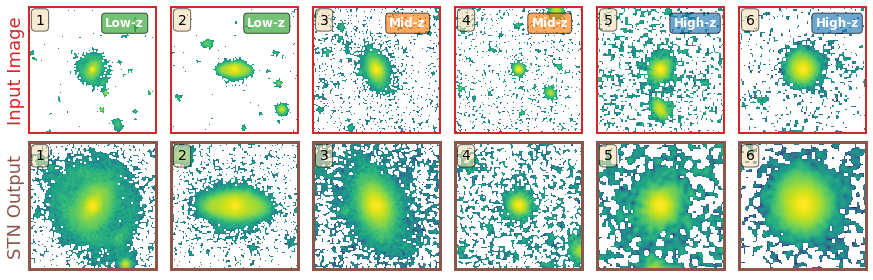}
    \caption{Examples of the transformation applied by the STN to two randomly selected galaxies from each redshift bin. The top row shows the input galaxy images, and the bottom row shows the corresponding output from the STN. The numbers in the top-left yellow boxes help correspond the output images to the input images. As can be seen, the STN learns to crop most secondary objects present in the input frame.}
    \label{fig:stn_random}
\end{figure*}

As can be seen from Rows 3, 4, and 5 of Figure \ref{fig:example_pred_dists}, \gampen{} can accurately predict morphological parameters even when the primary galaxy of interest occupies a small portion of the input cutout, and secondary objects are present in the input frame. This is primarily enabled by the upstream STN in \gampen{} which, during training, learns to apply an optimal amount of cropping to each input image. 

Figure \ref{fig:stn_random} shows examples of the transformations applied by the STN to randomly selected galaxies in all three redshift bins. As can be seen, the STN crops out most secondary galaxies present in the cutouts and helps the downstream CNN to focus on the galaxy of interest at the center.

\begin{figure*}[htb]
    \centering
    \includegraphics[width = 0.9\textwidth]{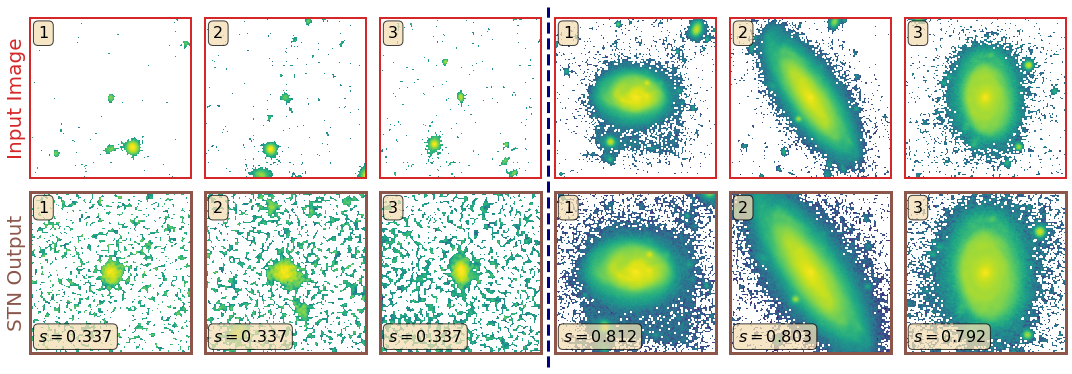}
    \caption{(\textit{Left}): Galaxies in the low-z bin  with the lowest values of $s$ (i.e., the most aggressive crops) (\textit{Right}): Galaxies in the low-z bin with the highest values of $s$ (i.e., the least aggressive crops). The $s$ parameter denotes the fraction of the input image that was retained in the STN output. As can be seen, the STN correctly learns to apply the most aggressive crops to small galaxies; and the least aggressive crops to large galaxies.}
    \label{fig:stn_min_max}
\end{figure*}

To further validate the performance of the STN, we measured the amount of cropping applied by the STN for all galaxies in the low-z bin. We chose the lowest redshift bin for this test as it has the largest range of galaxy radii among the different redshift bins. After that, we sorted all the processed images based on the amount of cropping applied to each input image. In Figure \ref{fig:stn_min_max}, we show example images from our dataset with extremely high and extremely low values of applied crops. The $s$ parameter shown in the lower left of each panel denotes what fraction of the input image was retained in the STN output---higher values of $s$ denote that a more significant fraction of the input image was retained in the output image produced by the STN  (i.e., minimal cropping). Figure \ref{fig:stn_min_max} demonstrates that (without us having to engineer this specifically), the STN correctly learns to apply the most aggressive crops to the smallest galaxies in our dataset, and the least aggressive crops to the largest galaxies

Thus, \gampen{}'s STN learns to systematically crop out secondary galaxies in the cutouts and focus on the galaxy of interest at the center of the cutout. At the same time, the STN also correctly applies minimal cropping to the largest galaxies, making sure the entirety of these galaxies remains in the frame. 

\subsection{Comparing \gampen{} predictions to GALFIT predictions} \label{sec:gampen_v_galfit}

\begin{figure*}[htb]
    \centering
    \includegraphics[width = 0.7\textwidth]{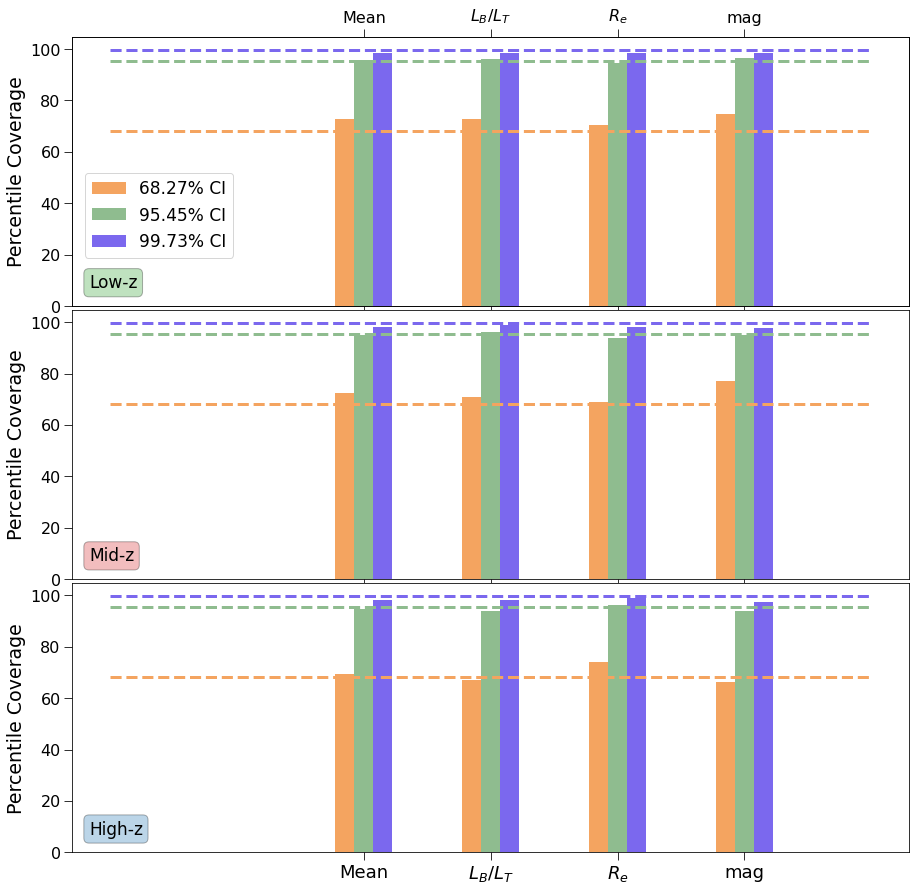}
    \caption{Percentile coverage probabilities achieved on the test set shown separately for each redshift bin. The leftmost set of bars in each panel shows the coverage probabilities when averaged over the three output parameters, and the right three sets of bars show the coverage probabilities for each parameter individually. The mean coverage probability never deviates by more than $4.5\%$ from the claimed confidence interval, and when considered for each parameter separately, the coverage probability never deviates by more than $8.7\%$. This demonstrates that \gampen{} produces well-calibrated accurate uncertainties.}
    \label{fig:cov_prob_all_z}
\end{figure*}

Out of the $60,000$ galaxies analyzed using GALFIT in \S \ref{sec:galfitting}, we use $80\%$ as the training and validation sets. We use the remaining $20\%$, which the trained \gampen{} frameworks have never seen, to evaluate the accuracy of the predicted parameters. We refer to this as the ``test set" henceforth.

In Figure \ref{fig:cov_prob_all_z}, we show the coverage probabilities achieved by \gampen{} on the test set. Note that in \S\,\ref{sec:fine_tuning_dropout}, we tuned the dropout rate using the validation set, whereas the values in Figure \ref{fig:cov_prob_all_z} are calculated on the test set. In the ideal situation, they would perfectly mirror the confidence levels; (e.g., $68.27\%$ of the time, the true value would lie within $68.27\%$ of the most probable volume of the predicted distribution). Clearly, the coverage probabilities achieved by \gampen{} are consistently close to the claimed confidence levels, both when averaged over the three output parameters, as well as for each parameter individually. The mean coverage probability never deviates by more than $4.5\%$ from the claimed confidence interval, and when considered for each parameter individually, the coverage probability never deviates by more than $8.7\%$ from the corresponding confidence interval. Additionally, we note that even for the case for which the coverage probabilities are most discrepant ($68\%$ flux confidence interval for the mid-z model), the uncertainties predicted by \gampen{} are in any case overestimates (i.e., conservative). If \gampen{} were used in a scenario that requires perfect alignment of coverage probabilities, users could employ techniques such as importance sampling \citep{importance_sampling} on the distributions predicted by \gampen{}. We note here that incorporating the covariances between the predicted parameters into our loss function was key to achieving simultaneous calibration of all three output variables. 

\begin{figure*}[htb]
    \centering
    \includegraphics[width = 0.9\textwidth]{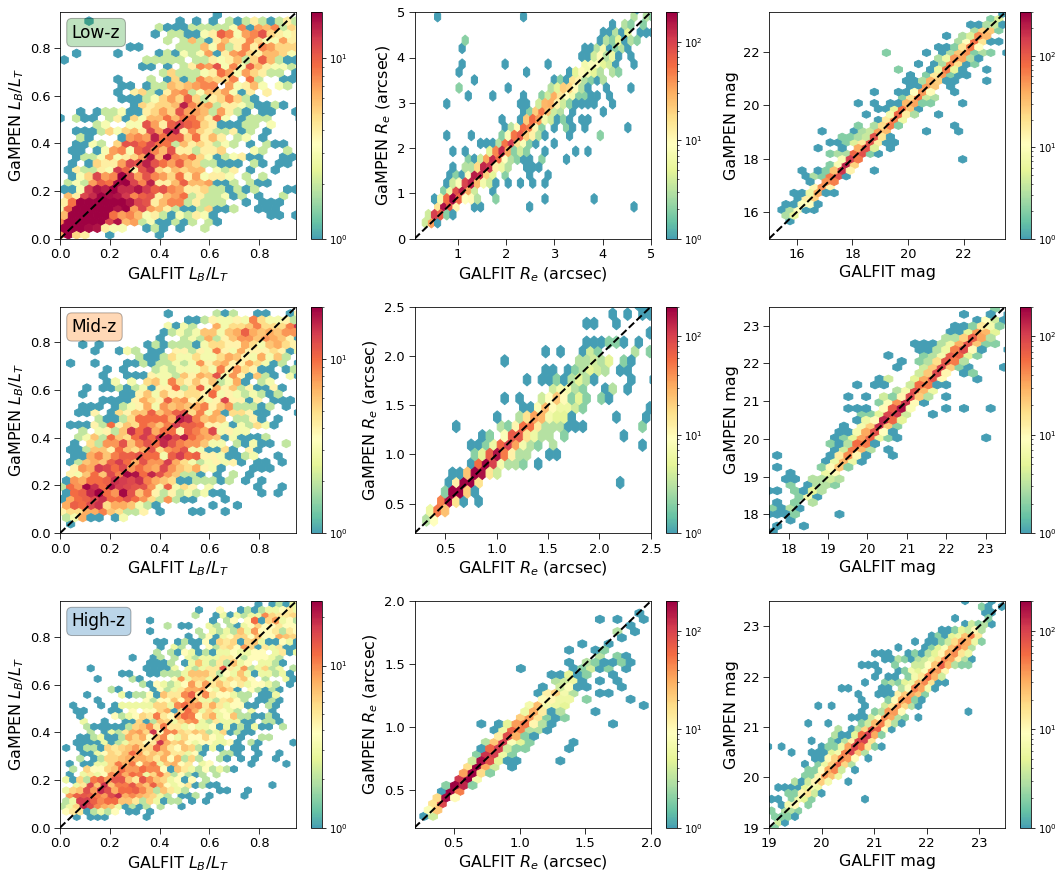}
    \caption{The most probable parameter values predicted by \gampen{} for all galaxies in the test set plotted against the values determined using GALFIT. Galaxies are plotted in hexagonal bins of roughly equal size, and the number of galaxies in each bin is represented according to the logarithmic colorbar to the right of each panel. The top, middle, and bottom rows show the results for the low-, mid-, and high-z bins, respectively. The dashed black $y=x$ line represents the line of equality. Across all three redshift bins, values predicted by \gampen{} closely mirror the values obtained using light-profile fitting.}
    \label{fig:pred_true_all_z}
\end{figure*}

Having shown above that the posteriors predicted by \gampen{} are well calibrated, we now investigate how close the modes (most probable values) of the predicted distributions are to the parameter values determined using light-profile fitting. Figure \ref{fig:pred_true_all_z} shows the most probable values predicted by \gampen{} for the test set plotted against the values determined using GALFIT in hexagonal bins of roughly equal size. The number of galaxies is represented according to the colorbar on the right. Note that we use a logarithmic colorbar to visualize even small clusters of galaxies in this plane, down to 1 galaxy/bin. Across all three redshift bins, a large majority of all the galaxies are clustered around the line of equality, showing that the most probable values of the distributions predicted by \gampen{} closely track the values obtained using light-profile fitting. The scatter obtained for $L_B/L_T$ is larger compared to the other two output parameters, and we explore that in more detail later in this section.

\begin{figure*}[htb]
    \centering
    \includegraphics[width = 0.9\textwidth]{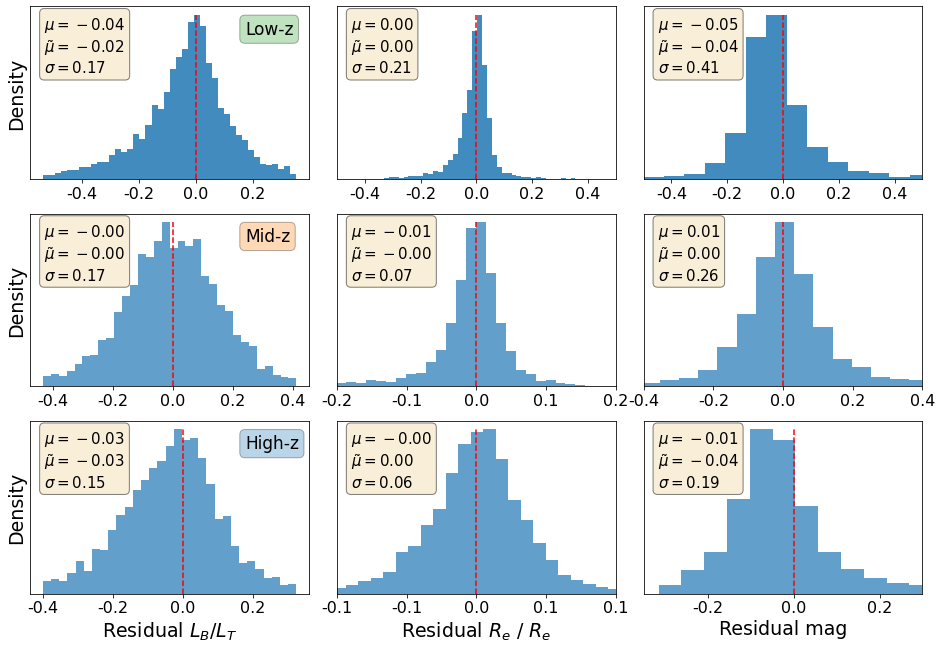}
    \caption{Distributions of residuals for all galaxies in the test set; specifically, the differences between the values predicted by \gampen{} and those obtained via light-profile fitting. The top, middle, and bottom rows show the results for the low-, mid-, and high-z bins, respectively. The boxes in the top-left corner of each panel show the mean ($\mu$), median ($\tilde{\mu}$), and standard deviation ($\sigma$) of each residual distribution. The $\sigma$ of each distribution identifies the typical disagreement for each parameter (e.g., for the low-z bin, in $68.27\%$ cases, the predicted magnitude is within $\pm0.41$ of the value determined by light profile fitting). The dashed red vertical line marks  $x=0$.}
    \label{fig:resi_all_z}
\end{figure*}

\textcolor{black}{In Figure \ref{fig:resi_all_z}, we show the residual distribution for \gampen{}'s output parameters in all three redshift bins. We define the residual for each parameter as the difference between the most probable value predicted by \gampen{} and the value determined using light profile fitting. The box in the upper left corner gives 
the mean ($\mu$), median ($\tilde{\mu}$), and standard deviation ($\sigma$) of each residual distribution. All nine distributions are normally distributed (verified using the Shapiro Wilk test), and have $\mu \sim \tilde{\mu} \sim 0$. The $\sigma$ of each distribution also identifies the typical disagreement for each parameter (e.g., for the low-z bin, in $68.27\%$ cases, the predicted $L_B/L_T$ value is within $\pm0.17$ of the value determined by light profile fitting). The residual $R_e$, when converted to physical units, correspond to typical disagreements of 0.32 arcsecs, 0.14 arcsecs, and 0.14 arcsecs for the low-, mid-, and high-z bins, respectively. The $L_B/L_T$ residuals are mostly constant across the three redshift bins, while the disagreement between \gampen{} and GALFIT predictions for $R_e$ and $F$ decrease slightly as we go from the low to the higher redshift bins. This could be driven by the fact that the HSC median seeing becomes better as we move from the g-band to the i-band (\gb{}-band: $0.\arcsec79$; \rb{}-band: $0.\arcsec75$; \ib{}-band: $0.\arcsec61$). Additionally, lower redshift galaxies have preferentially more resolved structural features (e.g., spiral arms), which are not accounted for in our disk + bulge GALFIT decomposition pipeline. This could lead to a higher disagreement between the GALFIT and \gampen{} predictions.}  

\begin{figure*}[htb]
    \centering
    \includegraphics[width = 0.9\textwidth]{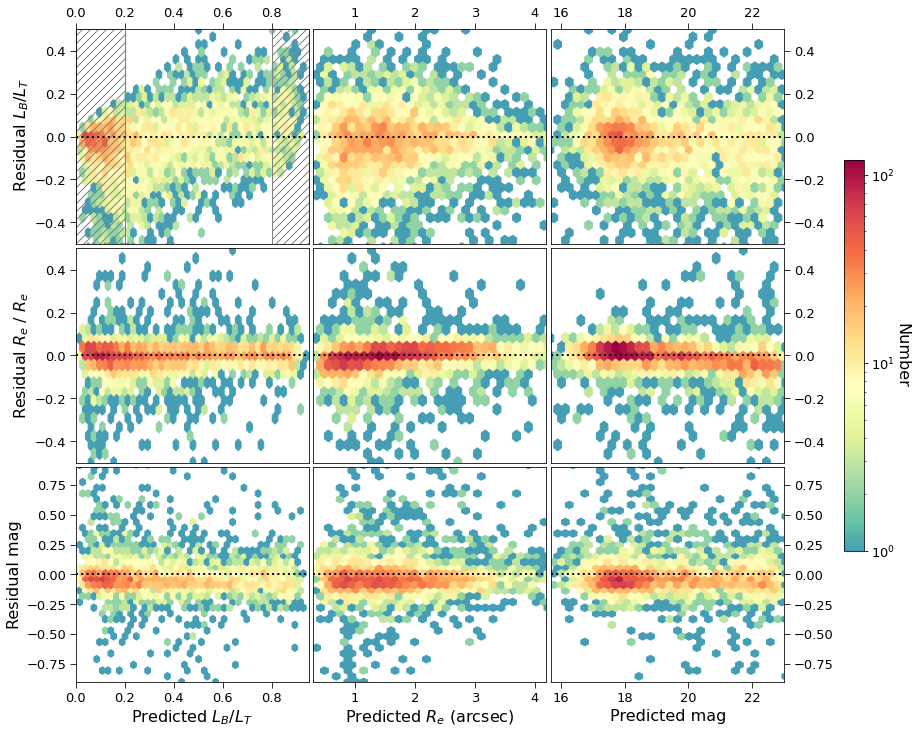}
    \caption{Residuals of the output parameters (difference between \gampen{} and GALFIT predictions) plotted against the values predicted by \gampen{} for all galaxies in the low-z test set. To make the y-axis dimensionless for all three parameters, we plot the fractional $R_e$ residuals instead of absolute values. This figure allows us to assign quality labels to \gampen{}'s predictions based on the output values (e.g., flagging regions of the parameter space with high levels of disagreement, as shown by the line-shaded region in the top-left panel). See \S\,\ref{sec:gampen_v_galfit} for details. The equivalent figures for the mid- and high-z bins are shown in Appendix \ref{ap:sec:2d_residuals}.}
    \label{fig:2d_res_low_z}
\end{figure*}

Although Figures \ref{fig:pred_true_all_z} and \ref{fig:resi_all_z} indicate the overall agreement between \gampen{} and GALFIT, they do not reveal the dependence of this agreement on location in the parameter space. This is critical information as this identifies regions of parameter space where \gampen{} agrees especially well or badly with light-profile fitting results, so that future users can flag the reliability of predictions in these regions or use results only from certain regions of the parameter space for specific scientific analyses. Figure \ref{fig:2d_res_low_z} shows the residuals for the three output parameters for the low-z bin plotted against the values predicted by \gampen{}. Note that in order to make the y-axis dimensionless for all three parameters, we plot the fractional $R_e$ residuals instead of absolute values. As in Figure \ref{fig:pred_true_all_z},  we have split the parameter space into hexagonal bins and used a logarithmic color scale to denote the number of galaxies in each bin. The trends between the residuals and different parameters are very similar across all three redshift bins. Thus, to keep the main text concise, we have shown the plot for the low-z bin here and shown the same plot for the mid- and high-z bins in Appendix \ref{ap:sec:2d_residuals}.

For most of the panels, the large majority of galaxies are clustered uniformly around the black dashed line, $y = 0$, which denotes the ideal case of perfectly recovered parameters (assuming the GALFIT parameters are correct). There are a few notable features on the top row, which depicts the $L_B/L_T$ residuals.  In the top left panel, the $L_B/L_T$ residuals are highest near the limits of $L_B/L_T$. We noticed the same effect when testing \gampen{} with simulated galaxies in \citet{gampen_software_paper}, and refer to this as the ``edge effect" 
For $L_B/L_T$ values near the edges (i.e., when the disk/bulge component completely dominates over the other component), precisely determining $L_B/L_T$ is challenging for \gampen{} --- in fact, this is difficult for any image analysis algorithm). In some of these cases, \gampen{} assigns almost the entirety of the light to the dominant component, resulting in the streaks seen at the edges of the figure. Poor structural parameter determination for galaxies with $L_B/L_T < 0.2$ and $L_B/L_T>0.8$ have also been independently observed in other studies using different algorithms \citep[e.g.,][]{euclid_morph, galapagos}

In order to mitigate this, \gampen{} users can choose to transform \gampen{}'s quantitative predictions in the region $0.2 \geq L_B/L_T \geq 0.8$ (demarcated by shaded lines in Figure \ref{fig:2d_res_low_z}) to qualitative values such as ``highly bulge-dominated" ($L_B/L_T \geq 0.8$) or ``highly disk-dominated" ($L_B/L_T \leq 0.2$). We followed a similar procedure in \citet{gampen_software_paper} and found the net accuracy of these labels to be $\gtrapprox95\%$.

\textcolor{black}{The top-middle panel of Figure \ref{fig:2d_res_low_z} also shows that $L_B/L_T$ residuals are higher for galaxies with smaller $R_e$. In other words, \gampen{} and GALFIT systematically disagree more for galaxies with smaller sizes---and this effect becomes more pronounced as the sizes become comparable to the seeing of the HSC-Wide Survey (\textit{g}-band: $0.79$ arcsec).} 

To comparatively evaluate the accuracy of \gampen{} and GALFIT specifically for smaller galaxies, we ran our GALFIT pipeline on a subset of simulated galaxies with $R_e \leq 2$ arcsec. Thereafter, we compared these results to the predictions made by \gampen{} for the same galaxies. As shown in Appendix \ref{ap:sec:gapemn_v_galfit}, \gampen{} outperforms GALFIT for these smaller simulated galaxies. This provides preliminary evidence that \gampen{}'s predictions on the smaller galaxies referred to in the previous paragraph are more accurate than those obtained using GALFIT. However, we would like to note that our simulated galaxies are semi-realistic and do not represent the full range of complexities present in real data. In a future publication, we will compare \gampen{} and GALFIT's performance on more realistic simulated galaxies generated using radiative transfer from hydrodynamical simulations.

\subsection{Inspecting the Predicted Uncertainties} \label{sec:uncertainties}

\begin{figure*}[htb]
    \centering
    \includegraphics[width = 0.9\textwidth]{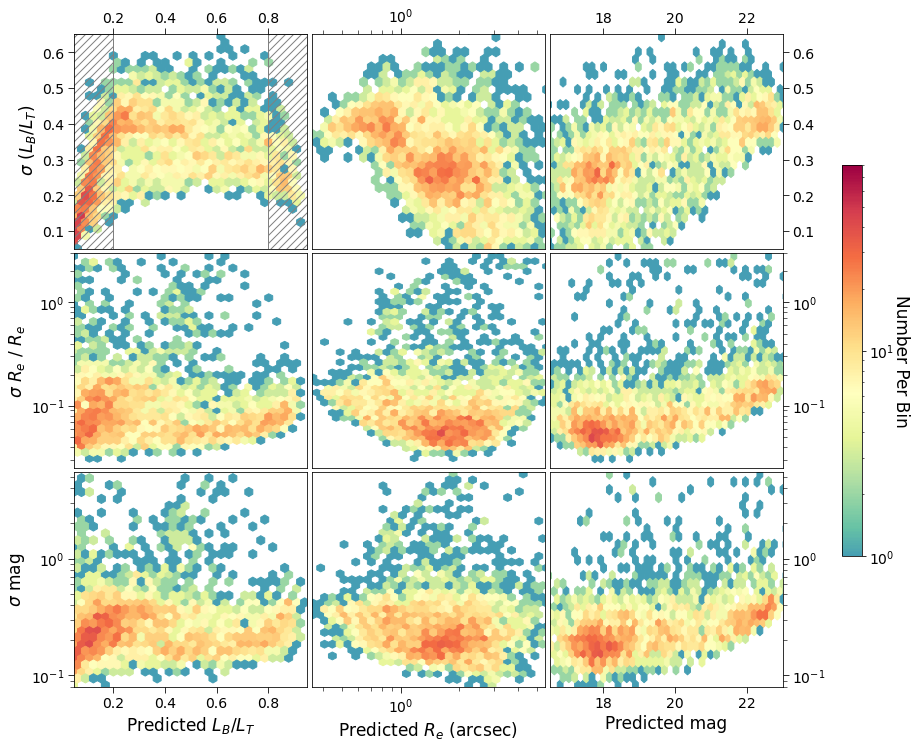}
    \caption{Uncertainties predicted by \gampen{} for each parameter plotted against the predicted values for the low-z test set. The $\sigma$ for each parameter is defined as the width of the $68.27\%$ confidence interval. Note that we plot fractional uncertainties for the radius in order to make the y-axis dimensionless for all three rows. The line-shaded region in the top-left panel shows the region where we recommend transforming quantitative $L_B/L_T$ predictions to qualitative labels (see \S \ref{sec:gampen_v_galfit} for details).}
    \label{fig:2d_uncer_low_z}
\end{figure*}

The primary advantage of a Bayesian ML framework like \gampen{} is its ability to predict the full posterior distributions of the output parameters instead of just point estimates. Thus, we would expect such a network to inherently produce wider distributions (i.e., larger uncertainties) in regions of the parameter space where residuals are higher. 

Figure \ref{fig:2d_uncer_low_z} shows the uncertainties for the three predicted parameters plotted against the predicted values for the low-z test set. We define the uncertainty predicted for each parameter as the width of the $68.27\%$ confidence interval (i.e., the parameter interval that contains $68.27\%$ of the most probable values of the predicted distribution; see Fig.\,\ref{fig:example_pred_dists}). The y-axis of the middle row has been normalized so that all three panels show dimensionless fractional uncertainties. The distributions of uncertainties look very similar across all redshifts; thus, we have shown the uncertainty distributions for the mid- and high-z bins in Appendix \ref{ap:sec:2d_residuals}.

In Figure \ref{fig:2d_res_low_z}, we saw that \gampen{}'s $L_B/L_T$ residuals are higher for lower values of $R_e$. Here, we see that \gampen{} accurately predicts higher $L_B/L_T$ uncertainties for lower values of $R_e$. This compensatory effect is what allows \gampen{} to achieve the calibrated coverage probabilities shown in Figure \ref{fig:cov_prob_all_z}. 

In the right column, we see that for all three parameters, the uncertainty in \gampen{}'s predictions increase for fainter galaxies. This is in line with what we expect and had seen in \citet{gampen_software_paper}---morphological parameters for fainter galaxies are more difficult to constrain compared to brighter galaxies and thus should have higher uncertainties. 

\textcolor{black}{The top left panel of Figure \ref{fig:2d_uncer_low_z} shows that \gampen{} is reasonably certain of its predicted 
bulge-to-total ratio across the full range of values but appears slightly more certain when $L_B/L_T \leq 0.2$ or $L_B/L_T \geq 0.8$. We had also seen the same effect in \citet{gampen_software_paper} with simulated galaxies. We found that the smaller uncertainties at the limits corresponded to the single-component galaxies, while for the double-component galaxies, the edge effect is less pronounced (see Figure 15 of \citet{gampen_software_paper}). Here, we are seeing the same effect--- \gampen{}'s uncertainties at the edges are systematically lower for galaxies that can be described completely by only a disk or bulge component. These lower uncertainties also contribute to the higher residuals near the edges of $L_B/L_T$ (as wider distributions would reduce the number of galaxies with high residuals at the edges). However, as can be seen from the top left panel of Figure \ref{fig:2d_uncer_low_z}, most of the galaxies with very low uncertainties lie in the region $0.2 \geq L_B/L_T \geq 0.8$, where we recommend transforming the quantitative $L_B/L_T$ predictions to qualitative labels, as outlined in \S \ref{sec:gampen_v_galfit}}.

The results shown in this section outline the primary advantage of using a Bayesian framework like \gampen{}--- even in situations where the network is not perfectly accurate, it can predict the right level of precision, allowing its predictions to be reliable and well-calibrated.

\subsection{Comparing \gampen{}'s Uncertainty Estimates to Other Algorithms}
\label{sec:uncer_comp}

\begin{figure*}[htb]
    \centering
    \includegraphics[width = 0.7\textwidth]{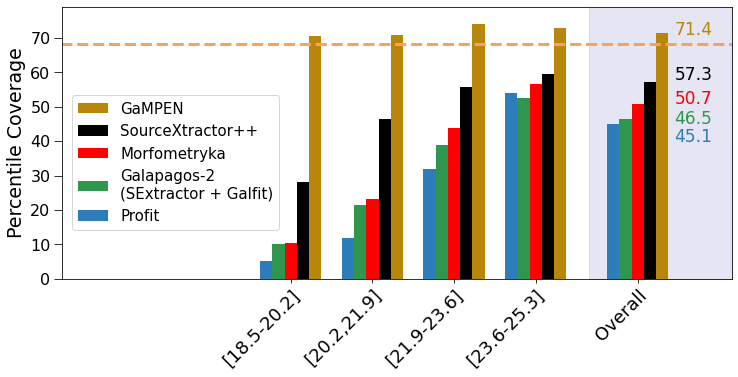}
    \caption{Percentile coverage probabilities for the $68.27\%$ confidence interval obtained by \gampen{} on our HSC sample compared to coverage probabilities obtained by various light-profile fitting algorithms on simulated Euclid data 
    (from \citealp{euclid_morph}). The rightmost set of bars shows the values calculated on the entire dataset, while the other sets display values calculated on sub-samples of galaxies with specific magnitude ranges (AB mag, shown on the x-axis). 
    Compared to light-profile fitting tools, the uncertainties predicted by \gampen{} are better calibrated by $\sim15-25\%$ overall and by as much as $\sim60\%$ for the brightest galaxies.}
    \label{fig:cov_prob_comp_euclud}
\end{figure*}

As \S \ref{sec:gampen_v_galfit} and \S\ref{sec:uncertainties} demonstrate, \gampen{} predicts well-calibrated uncertainties on our HSC data. Previous studies \citep[e.g.,][]{haussler_07} have shown that analytical estimates of errors from traditional morphological analysis tools like GALFIT or GIM2D \citep{gim2d} are smaller than the true uncertainties by $\geq70\%$ for most galaxies.

Recently, \citet{euclid_morph} reported coverage probabilities obtained by four different light-profile fitting tools---Galapagos-2 \citep{galapagos}, Morfometryka \citep{morfometryka}, ProFit \citep{profit}, and SourceXtractor++\citep{srcx++}---on simulated Euclid data. The Euclid sample consisted of $\sim1.5$ million galaxies ranging from $I_E\sim15$ to $I_E\sim30$, simulated at $0.\arcsec1$ /pixel. The simulations included analytic \sersic{} profiles with one and two components, as well as more realistic galaxies generated with neural networks. 
As we did not have access to the simulated Euclid dataset, we could not test \gampen{}'s performance on the same data. 
Instead, we compared \gampen{}'s coverage probabilities for the HSC data set to those reported for the Euclid simulations. 
Although the 
latter is significantly different from our HSC sample, coverage probabilities reflect the ability of the predicted uncertainty to capture the true uncertainty and are not necessarily correlated with accuracy (which often varies across different data sets). Moreover, \gampen{}'s predicted uncertainties can be tuned for specific data sets, as shown in Figure \ref{fig:dropout_calibration}, which should only improve the \gampen{} outcome. Therefore, the results presented by \citet{euclid_morph} allow us to perform a preliminary comparison of \gampen{}'s uncertainty prediction to that of other algorithms.

Figure \ref{fig:cov_prob_comp_euclud} shows the $68.27\%$ coverage probabilities achieved by \gampen{} on the HSC data compared to values for the four light profile fitting codes on the simulated Euclid dataset 
(averaging over the different structural parameters). 
When considering all galaxies, \gampen{}'s uncertainties are at least $\sim15-25\%$ better calibrated than the other algorithms. 
The differences are much larger for brighter galaxies,
suggesting that the uncertainties predicted by these algorithms depend primarily on the flux of the object. In contrast, \gampen{}'s uncertainty predictions remain well-calibrated throughout and are better by as much as $\sim60\%$ for the brightest galaxies. The severe under-prediction of uncertainties seems to hold true even for Bayesian codes like ProFit. 
It is likely that 
\gampen{}'s robust implementation of aleatoric and epistemic uncertainties, along with a carefully selected transfer learning set spanning the entire magnitude range (see Fig. \ref{fig:tl_para_dist}), allows it to predict well-calibrated uncertainties across a wide range of magnitudes.

\section{Comparing Our Predictions to Other Catalogs} \label{sec:compare}

\begin{figure*}[htb]
    \centering
    \includegraphics[width = 0.9\textwidth]{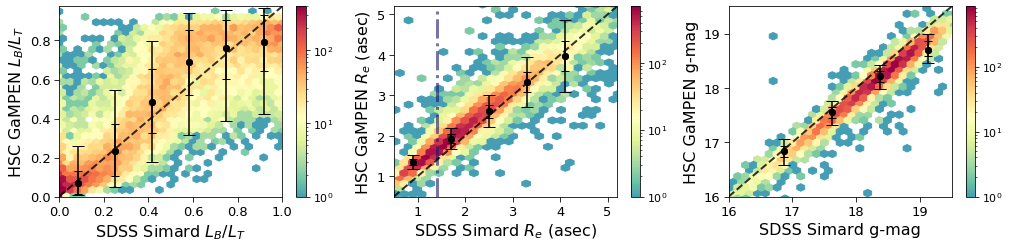}
    \caption{\gampen{} predictions plotted against values estimated by \citet{simard_11} for a cross-matched sample of $\sim20,000$ galaxies with $z < 0.2$ and $m < 19$. The density of points in each histogram is represented according to the logarithmic colorbar on the right. The black dots show the median y values in bins of equal width along the x-axis, with the error bars depicting the average $68.27\%$ and $95.45\%$ confidence intervals predicted by \gampen{} in that bin. The dash-dotted vertical line in the middle panel shows the median SDSS \gb{}-band seeing.}
    \label{fig:gampen_v_simard}
\end{figure*}

\begin{figure*}[htb]
    \centering
    \includegraphics[width = 0.9\textwidth]{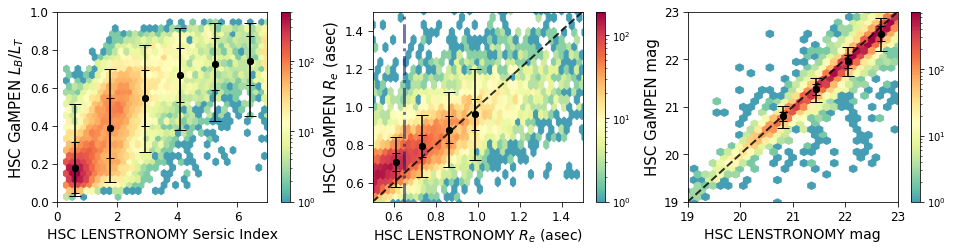}
    \caption{\gampen{} predictions plotted against values estimated by \citet{hsc_sersic} for a cross-matched sample of $\sim200,000$ galaxies with $0.5 < z \leq 0.75$ and $m \leq 23$. Similar to Figure \ref{fig:gampen_v_simard}, we show the median y-values and associated error bars in bins of equal width along the x-axis. The dash-dotted vertical line in the middle panel shows the median HSC \ib{}-band seeing.}
    \label{fig:gampen_v_lenstronomy}
\end{figure*}

In \S \ref{sec:gampen_v_galfit}, we compared \gampen{}'s predictions to the values determined using our light-profile fitting pipeline. In order to further assess the reliability of \gampen{}'s predictions, we now compare our predictions to two other morphological catalogs. 

In Figure \ref{fig:gampen_v_simard}, we compare \gampen{}'s predictions to the fits of \citet{simard_11}. Using an angular cross-match diameter of $0.\arcsec15$, we cross-matched our entire sample to that of \citet{simard_11} to obtain an overlapping sample of $\sim20,000$ galaxies. A large majority of these galaxies have $m < 19.5$ and $z < 0.2$. We have included error bars in this figure to show the typical width of predicted distributions for different regions of the parameter space. We binned the x-axis of each parameter into bins of equal width and plotted the median y-value in each bin as a point, with the error bars showing the average $68.27\%$ and $95.45\%$ confidence intervals for all galaxies in that bin. 

\textcolor{black}{Despite the differences in SDSS and HSC imaging quality (with regards to pixel-scale and seeing), our results largely agree with those of \citet{simard_11} within the ranges of predicted uncertainties. Using Spearman's rank correlation test, we obtain correlation coefficients of 0.87, 0.96, and 0.98 for $L_B/L_T$, $R_e$, and $F$, respectively. It is also interesting to note that the widths of the error bars account for almost the entire scatter in the distribution of points, showing the robustness of \gampen{}'s uncertainty estimates.}

In the $L_B/L_T$ panel, there is a cluster of points near $L_B/L_T = 0.8$, and this is due to the edge-effect described in \S \ref{sec:gampen_v_galfit}. \gampen{} and \citet{simard_11}'s $R_e$ predictions become discrepant (even when accounting for uncertainty) for $R_e \sim 1\arcsec$. This is not unexpected given that these $R_e$ values are almost $0.\arcsec4$ smaller than the median SDSS seeing, depicted by the dash-dotted vertical line in the middle panel of Figure \ref{fig:gampen_v_simard}. The slight offset seen in the magnitude estimates is due to the differences in SDSS and HSC imaging quality, data-reduction pipeline, and filters \citep{hsc_filters}, given that this discrepancy disappears when we compare our results to another catalog that uses HSC imaging (see below). Note that some of the scatter in Figure 22 can also be attributed to the fact that \citet{simard_11} uses a morphology-determination pipeline that is significantly different from \gampen{}.

\textcolor{black}{In Figure \ref{fig:gampen_v_lenstronomy}, we compare our results to that of \citet{hsc_sersic}\footnote{Catalog obtained from John D. Silverman, private communication.}, wherein the authors fitted single \sersic{} light-profiles to $1.5\times10^6$ HSC \ib{}-band galaxies using \lenstronomy{} \citep{Birrer2018Lenstronomy:Package}, a multipurpose open-source gravitational lens modeling Python package. Following Figure \ref{fig:gampen_v_simard}, we also show mean error bars in different bins of the fitted parameter in this figure. To compare results in the same band, we cross-matched our high-z sample with that of \citet{hsc_sersic} to obtain a sample of $\sim200,000$ galaxies with $0.50 < z \leq 0.75$ and \ib{} $\leq23$. Note that the magnitude discrepancy present in Figure \ref{fig:gampen_v_simard} now disappears. \gampen{} and \lenstronomy{} radius measurements are also in agreement within the limit of uncertainties, with the mean trend deviating slightly from the $y=x$ diagonal at $R_e$ values lower than the HSC median seeing, depicted by the dash-dotted vertical line in Figure \ref{fig:gampen_v_lenstronomy}. It is also important to note that while we define $R_e$ as the radius that contains $50\%$ of the total light of the combined bulge + disk profile, \citet{hsc_sersic} defines it as the semi-major axis of the ellipse that contains half of the total flux of the best-fitting \sersic{} model. The correlation coefficients for $R_e$ and magnitude are 0.87 and 0.96, respectively.} 

\textcolor{black}{Since \citet{hsc_sersic} used single-component fits, we cannot compare our $L_B/L_T$ predictions with their catalog. However, the left panel of Figure \ref{fig:gampen_v_lenstronomy} shows the correlation between fitted \sersic{} index ($n$) and measured bulge-to-total light ratio and can be used empirically to convert one measure into another. We find that, in line with expectations, higher \sersic{} indices generally correspond to higher values of $L_B/L_T$. A large majority of galaxies with $n \leq 1.5$ have most of their light in the disk component. Galaxies with $ n \geq 3$ have a large fraction of their light in the bulge component, although this fraction may vary from values as low as $40\%$ to $95\%$. Note that these trends largely agree with what was reported by \citet{simmons_08}.}

\section{Conclusions \& Discussion} \label{sec:conclusions}
In this paper, we used \gampen{}, a Bayesian machine learning framework, to estimate morphological parameters ($L_B/L_T$, $R_e$, $F$) and associated uncertainties for $\sim8$ million galaxies in the HSC Wide survey with $z \leq 0.75$ and $m \leq 23$. Our catalog is one of the largest morphological catalogs and is the first publicly available structural parameter catalog for HSC galaxies. It provides an order of magnitude more galaxies compared to the current state-of-the-art disk+bulge decomposition catalog of \citet{simard_11} while probing four magnitudes deeper and having a higher redshift threshold. This represents an important step forward in our capability to quantify the shapes and sizes of galaxies and uncertainties therein.

We also demonstrated that by first training on simulations of galaxies and then utilizing transfer learning using real data, we are able to train \gampen{} using $<1\%$ of our total dataset for training. This is an important demonstration that ML frameworks can be used to measure galaxy properties in new surveys, which do not have already-classified large training sets readily available. Our implemented two-step process provides a new framework that can be easily used for upcoming large imaging surveys like the Vera Rubin Observatory Legacy Survey of Space and Time, Euclid, and the Nancy Grace Roman Space Telescope. 

\textcolor{black}{We showed that \gampen{}'s STN is adept at automatically cropping input frames and successfully removes secondary objects present in the frame for most input cutouts. Note that the trained STN framework can be detached from the rest of \gampen{}, and can be used as a pre-processing step in any image analysis pipeline.}

\textcolor{black}{By comparing \gampen{}'s predictions to values obtained using light-profile fitting, we demonstrated that the posteriors predicted by \gampen{} are well-calibrated and \gampen{} can accurately recover morphological parameters. We note that the full computation of Bayesian posteriors in \gampen{} represents a significant improvement over estimates of errors from traditional morphological analysis tools like GALFIT, Galapagos, or ProFit. We demonstrated that while the uncertainties predicted by \gampen{} are well calibrated with $\lesssim5\%$ deviation across a wide range of magnitudes, traditional light-profile fitting algorithms underestimate uncertainties by $\sim15-60\%$ depending on the flux of the galaxy being analyzed. These well-calibrated uncertainties will allow us to use \gampen{} for the derivation of robust scaling relations  \citep[e.g.,][]{Bernardi2013TheProfile, vanderWel20143D-HST+CANDELS:3} as well as for tests of theoretical models using morphology \citep[e.g.,][]{Schawinski2014TheGalaxies}.}

\textcolor{black}{\gampen{}'s residuals increase for smaller galaxies, but \gampen{} correctly accounts for that by predicting correspondingly higher uncertainties for these galaxies. \gampen{}'s $L_B/L_T$ residuals are also high when the bulge or disk component completely dominates over the other component. \gampen{}'s quantitative $L_B/L_T$ predictions for these galaxies ($0.2 \geq L_B/L_T \geq 0.8$) can be transformed into highly accurate qualitative labels with $\gtrsim95\%$ accuracy. We did not detect a decline in \gampen{}'s performance for fainter galaxies, and \gampen{}'s residual patterns were also fairly consistent across all three redshift bins used in this study.}

\textcolor{black}{In order to assess the reliability of our catalog using a completely independent analysis, we compared \gampen{}'s predictions to those of \citet{simard_11} and \citet{hsc_sersic}. Within the limit of uncertainties predicted by \gampen{}, our results agree well with these two catalogs. We noticed a slight discrepancy in the flux values determined using SDSS and HSC imaging, which can be attributed to the differences in imaging quality, data-reduction pipeline, and filters between the two surveys. Comparing our $L_B/L_T$ predictions to \citet{hsc_sersic}'s measured \sersic{} indices also allowed us to study the correlation between these two parameters, and this can also be used empirically to switch between these two parameters. We found that although galaxies with $n\geq3$ have a large fraction of their light in the bulge component, this fraction can be anywhere between $40\%$ to $95\%$.}

Similar to other previous structural parameter catalogs \citep[e.g.,][]{simard_11,tarsitano_18}, we did not explicitly exclude merging galaxies from this catalog. We are currently working to incorporate a prediction flag within \gampen{} that will flag merging/irregular galaxies and highly blended sources so that they can be analyzed separately --- we will demonstrate this in future publications. However, we would like to note that for the redshift ranges and magnitudes considered in this study, blending and mergers only constitute a limited fraction of the total sample. Using the \texttt{m\_(g|r|i)\_blendedness\_flag} and \texttt{m\_(g|r|i)\_blendedness\_abs\_flux} parameters available in HSC PDR2, we estimate that $\sim10\%$, $\sim4\%$, and $\sim3\%$ of galaxies in the low-z, mid-z, and high-z redshift bins, respectively, have nearby sources within the parent object footprint that can affect structural parameter measurements. For an extended description of these flags, we refer the interested reader to \citet{hsc_pipeline} and note that users can choose to ignore/exclude these galaxies from their analysis by setting different values/thresholds for these two parameters. 

\textcolor{black}{With this work, we are publicly releasing: a) \gampen{}'s source code and trained models, along with documentation and tutorials; b) A catalog of morphological parameters for the $\sim8$ million galaxies in our sample along with robust estimates of uncertainties; c) The full posterior distributions for all $\sim8$ million galaxies. All elements of the public data release are summarized in Appendix \ref{sec:ap:data_access}.}

\textcolor{black}{Although we used \gampen{} here to predict $L_B/L_T$, $R_e$, and $F$, it can be used to predict any morphological parameter when trained appropriately. Additionally, although \gampen{} was used here on single-band images, we have tested that both the STN and CNN modules in \gampen{} can handle an arbitrary number of channels, with each channel being a different band. We defer a detailed evaluation of \gampen{}'s performance on multi-band images for future work. Additionally, \gampen{} can also be used to morphologically analyze galaxies from other ground and space-based observatories. However, in order to apply \gampen{} on these data sets, one would need to perform appropriate transfer learning using data from the target dataset.}

\textcolor{black}{Finally, to give readers an estimate of \gampen{}'s run-time, we note that once trained, it takes \gampen{} $\sim1$ millisecond on a GPU and $\sim150$ milliseconds on a CPU to perform a single forward pass on an input galaxy. These numbers, of course, change slightly based on the specifics of the hardware being used. However, as these numbers show, even with access to $\sim1000$ CPUs or $\sim10$ GPUs, \gampen{} can estimate full Bayesian posteriors for millions of galaxies in just a few days. Therefore, \gampen{} is fully ready for the large samples expected soon from Rubin-LSST and Euclid. Determinations of structural parameters along with robust uncertainties for these samples will allow us to characterize both morphologies as well as other relevant properties traced by morphology (e.g., merger history) as a function of cosmic time, mass, and the environment with unmatched statistical significance.}

\section*{acknowledgments}
This material is based upon work supported by the National Science Foundation under Grant No. 1715512.

CMU and AG would like to acknowledge support from the National Aeronautics and Space Administration via ADAP Grant 80NSSC18K0418. 

AG would like to acknowledge the support received from the Yale Graduate School of Arts \& Sciences through the Dean's Emerging Scholars Research Award.

AG would like to acknowledge computing grants received through the Amazon Cloud Credits for Research Program and the Yale Center for Research Computing (YCRC) Research Credits Program. AG would also like to thank the Yale Center for Research Computing and Yale Information Technology Services staff members and scientists, especially Robert Bjorson and Craig Henry, for their guidance and assistance in the vast amount of computation required for this project that was performed on Yale's Grace computing cluster and the Yale Astronomy compute nodes. 

We would like to thank John D. Silverman and Lalitwadee Kawinwanichakij for making the \citet{hsc_sersic} catalog available to us. AG would like to thank Tim Miller and Imad Pasha for helpful discussions.

PN gratefully acknowledges support at the Black Hole Initiative (BHI) at Harvard as an external PI with grants from the Gordon and Betty Moore Foundation and the John Templeton Foundation.

ET acknowledges support from ANID through Millennium Science Initiative Program - NCN19\_058, CATA-BASAL ACE210002 and FB210003, and FONDECYT Regular 1190818 and 1200495.

The Hyper Suprime-Cam (HSC) collaboration includes the astronomical communities of Japan and Taiwan, and Princeton University. The HSC instrumentation and software were developed by the National Astronomical Observatory of Japan (NAOJ), the Kavli Institute for the Physics and Mathematics of the Universe (Kavli IPMU), the University of Tokyo, the High Energy Accelerator Research Organization (KEK), the Academia Sinica Institute for Astronomy and Astrophysics in Taiwan (ASIAA), and Princeton University. Funding was contributed by the FIRST program from Japanese Cabinet Office, the Ministry of Education, Culture, Sports, Science and Technology (MEXT), the Japan Society for the Promotion of Science (JSPS), Japan Science and Technology Agency (JST), the Toray Science Foundation, NAOJ, Kavli IPMU, KEK, ASIAA, and Princeton University. 

This paper makes use of software developed for the Large Synoptic Survey Telescope. We thank the LSST Project for making their code available as free software at  \href{http://dm.lsst.org}{http://dm.lsst.org}.

The Pan-STARRS1 Surveys (PS1) have been made possible through contributions of the Institute for Astronomy, the University of Hawaii, the Pan-STARRS Project Office, the Max-Planck Society and its participating institutes, the Max Planck Institute for Astronomy, Heidelberg and the Max Planck Institute for Extraterrestrial Physics, Garching, The Johns Hopkins University, Durham University, the University of Edinburgh, Queen’s University Belfast, the Harvard-Smithsonian Center for Astrophysics, the Las Cumbres Observatory Global Telescope Network Incorporated, the National Central University of Taiwan, the Space Telescope Science Institute, the National Aeronautics and Space Administration under Grant No. NNX08AR22G issued through the Planetary Science Division of the NASA Science Mission Directorate, the National Science Foundation under Grant No. AST-1238877, the University of Maryland, and Eotvos Lorand University (ELTE) and the Los Alamos National Laboratory.

Based, in part, on data collected at the Subaru Telescope and retrieved from the HSC data archive system, which is operated by Subaru Telescope and Astronomy Data Center at National Astronomical Observatory of Japan.

\clearpage

\appendix

\section{Data Access}\label{sec:ap:data_access}

GaMPEN's source code is being publicly released along with the publication of this article. Along with the source code, we are also releasing documentation, tutorials, the trained models, the entire catalog of morphological predictions as well as the estimated PDFs for different morphological parameters of the $\sim8$ million galaxies in our sample. 

\vspace{10pt}
Links to the various components of the public data release mentioned above can be accessed at:-
\begin{itemize}
    \item Summary -- \href{http://gampen.ghosharitra.com/}{\url{http://gampen.ghosharitra.com/}}
    \item Summary (Mirror of Above) --\href{http://www.astro.yale.edu/aghosh/gampen.html}{\url{http://www.astro.yale.edu/aghosh/gampen.html}}
    \item Source Code --
    \href{https://github.com/aritraghsh09/GaMPEN}{\url{https://github.com/aritraghsh09/GaMPEN}}
    \item Documentation \& Tutorials -- 
    \href{https://gampen.readthedocs.io/en/latest/}{\url{https://gampen.readthedocs.io/en/latest/}}
\end{itemize}

We caution users to carefully read the Public Data Release Handbook available at \href{https://gampen.readthedocs.io/en/latest/Public_data.html}{\url{https://gampen.readthedocs.io/en/latest/Public_data.html}} to understand various aspects of the data release before using the morphological catalogs produced as a part of this paper. 

Along with the source code and trained models, this public data release also includes the various parameters obtained using the light profile fitting pipeline described in \S \ref{sec:galfitting}.

Since \gampen{} is a living code repository, which we expect to keep changing with time, we have created a ``frozen" version of the code at the time of writing this article. This version is tagged as release \texttt{v0.1.0} in the above-mentioned Github repository and can also be referred to as \citet{gampen_first_release}.

To ensure the long-term availability of the public data release's most critical components, we have also uploaded the catalog files and trained models to Zenodo at doi:\href{https://zenodo.org/records/8067382}{10.5281/zenodo.8067382}. We recommend users use the Zenodo files in case any of the links above appear broken.

\section{Additional Details About Data}\label{ap:data}

\begin{figure}[htb]
    \centering
    \includegraphics[width = 0.47\textwidth]{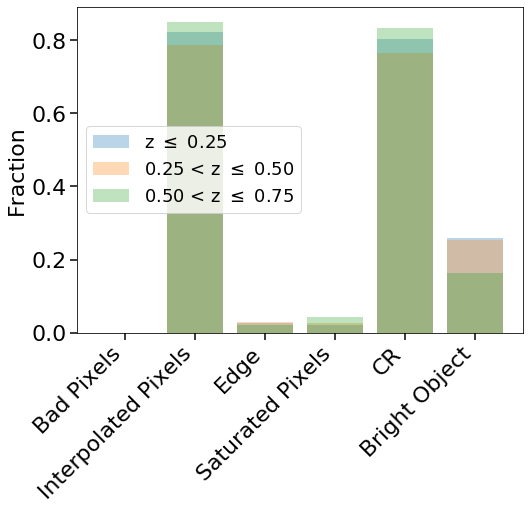}
    \caption{We exclude $\sim20\%$ of our downloaded galaxies due to different flags being triggered. The distribution of various flags that contribute to galaxies being excluded from our sample is shown in this figure. As can be seen, the large majority of exclusions are due to cosmic-ray hits (and hence, interpolated pixels). }
    \label{fig:flag_distr}
\end{figure}

As outlined in \S\ref{sec:data}, we used the \texttt{cleanflags\_any} parameter available as part of HSC PDR2 to exclude objects flagged to have any significant imaging issues by the HSC pipeline. The various triggers which contribute to the above flag, as well as their prevalence among the galaxies which we excluded from the analysis, are shown in Figure \ref{fig:flag_distr}. As can be seen, $\sim80\%$ of the triggers are caused by cosmic ray hits/interpolated pixels.

In addition, the full SQL queries used to download the low-, mid-, and high-z data are shown in Listings \ref{code:sql_low_z}, \ref{code:sql_mid_z}, and \ref{code:sql_high_z}. Note that after downloading the data using these queries, we further excluded data based on the flags referred to in the previous paragraph, as well as the quality of photometric redshift estimates. For an extended description, please 
refer to \S \ref{sec:data}. As noted in \S \ref{sec:conclusions}, users may additionally choose to use the various \texttt{blendedness} flags available in HSC PDR2 to further exclude merging/blended galaxies. 

\begin{listing}[H]
\begin{minted}[
frame=lines,
framesep=2mm,
baselinestretch=1.2,
fontsize=\footnotesize,
bgcolor=LightGray,
linenos,
breaklines,
]{sql}
SELECT
    object_id
      , ra
      , dec
FROM
    pdr2_wide.smallcat
    LEFT JOIN pdr2_wide.specz USING (object_id)
    LEFT JOIN pdr2_wide.photoz_mizuki USING (object_id)
    WHERE
    ((photoz_best > -1 AND photoz_best <= 0.25) OR (specz_redshift > -1 AND specz_redshift <= 0.25))
    AND (g_kronflux_mag < 23 OR g_cmodel_mag < 23)
    AND g_extendedness_value = 1
;
\end{minted}
\caption{Low-z Sample SQL query}
\label{code:sql_low_z}
\end{listing}

\begin{listing}[H]
\begin{minted}[
frame=lines,
framesep=2mm,
baselinestretch=1.2,
fontsize=\footnotesize,
bgcolor=LightGray,
linenos,
breaklines,
]{sql}

SELECT
    object_id
      , ra
      , dec
FROM
    pdr2_wide.smallcat
    LEFT JOIN pdr2_wide.specz USING (object_id)
    LEFT JOIN pdr2_wide.photoz_mizuki USING (object_id)
    WHERE
    ((photoz_best > 0.25 AND photoz_best <= 0.50) OR (specz_redshift > 0.25 AND specz_redshift <= 0.50))
    AND (r_kronflux_mag < 23 OR r_cmodel_mag < 23)
    AND r_extendedness_value = 1
;
\end{minted}
\caption{Mid-z Sample SQL query}
\label{code:sql_mid_z}
\end{listing}

\begin{listing}[H]
\begin{minted}[
frame=lines,
framesep=2mm,
baselinestretch=1.2,
fontsize=\footnotesize,
bgcolor=LightGray,
linenos,
breaklines,
]{sql}

SELECT
    object_id
      , ra
      , dec  
FROM
    pdr2_wide.smallcat
    LEFT JOIN pdr2_wide.specz USING (object_id)
    LEFT JOIN pdr2_wide.photoz_mizuki USING (object_id)
    WHERE
    ((photoz_best > 0.50 AND photoz_best <= 0.75) OR (specz_redshift > 0.50 AND specz_redshift <= 0.75))
    AND (i_kronflux_mag < 23 OR i_cmodel_mag < 23)
    AND i_extendedness_value = 1
;
\end{minted}
\caption{High-z Sample SQL query}
\label{code:sql_high_z}
\end{listing}

\section{Additional Details About \gampen{}}\label{ap:gampen}

\begin{figure*}[htb]
    \centering
    \includegraphics[width
    =\textwidth]{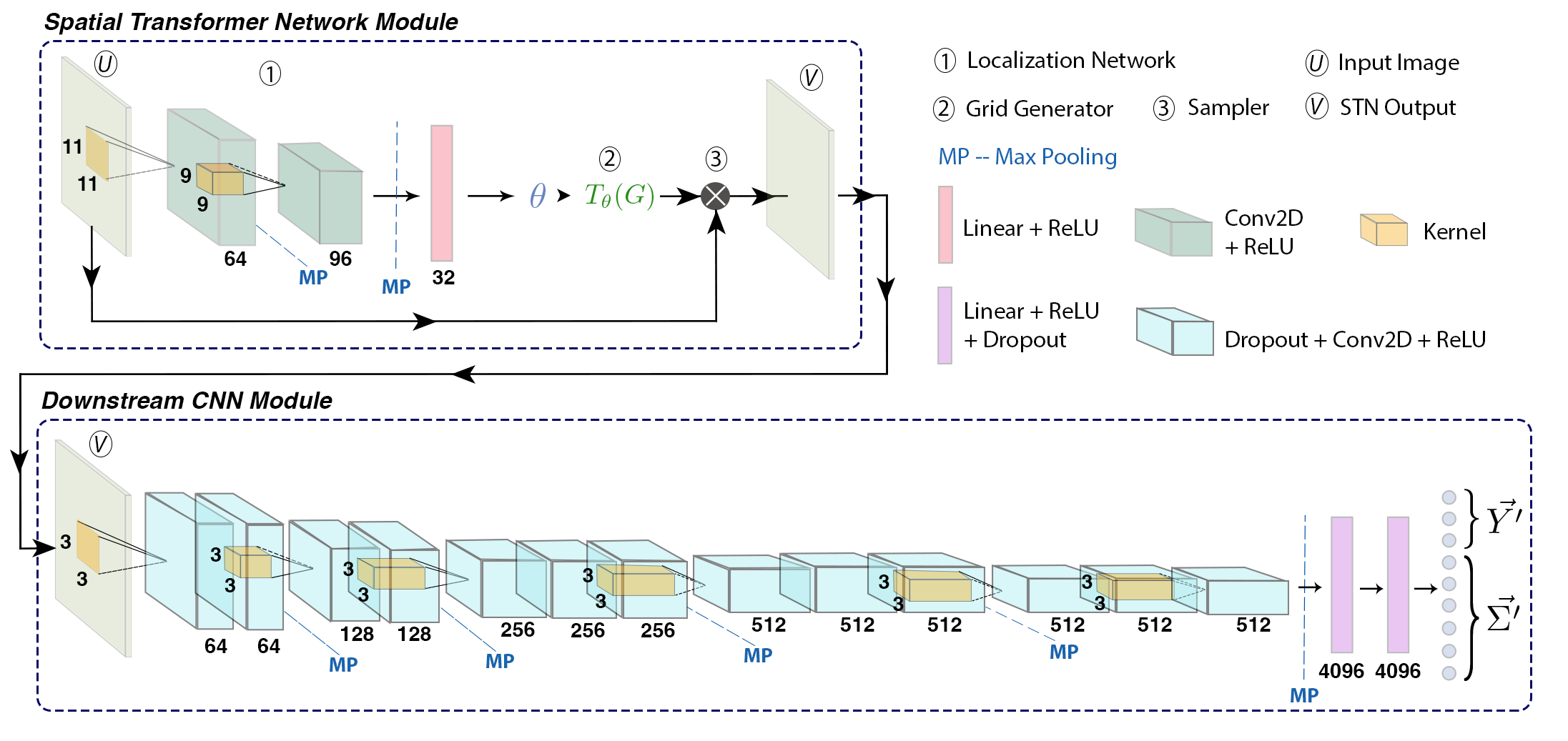}
    \caption{A schematic diagram of the Galaxy Morphology Posterior Estimation Network. \gampen's architecture consists of a downstream CNN module preceded by an upstream STN module. The CNN module empowers \gampen{} to estimate posterior distributions of galaxy morphology parameters. The upstream STN module trains without any extra supervision and learns to apply appropriate cropping transformations to the input image before passing it on to the CNN (for more details about these modules, see \S \ref{sec:gampen}).
    The numbers below each layer refer to the number of filters/neurons in each layer. The yellow boxes inside the convolutional layers show the kernel and the number beside it refers to the corresponding kernel size. Only one kernel is shown per set of convolutional layers; all other layers in the set have kernels of the same size. Conv2D and ReLU refer to Convolutional Layers and Rectified Linear Units, respectively.}
    \label{fig:gampen_schematic}
\end{figure*}

To provide readers a visual understanding of how \gampen{}'s different architectural components are organized, Figure \ref{fig:gampen_schematic} shows a schematic diagram outlining the structure of both the STN and CNN in \gampen{}. For complete details on individual layers in \gampen{}, please refer to \citet{gampen_software_paper}.

\section{Additional Details About Trained \gampen{} Models}\label{ap:sec:trained_models}

As noted in \S \ref{sec:method}, the training procedure of \gampen{} involves the tuning of various hyper-parameters (e.g., learning rate, batch size, etc.). These hyper-parameters are chosen based on the combination of the values that result in the best performance on the validation data set. The final chosen hyper-parameters for both the simulation trained \gampen{} models, as well as those fine-tuned on real data, are shown in Table \ref{tab:hyper_para}. Please refer to \S \ref{sec:method} for more details on how we train these models.

\begin{deluxetable*}{c|cccccc}[htbp]
\tablecaption{Tuned Values of Various Hyper-parameters  \label{tab:hyper_para}}
\tablecolumns{6}
\tablehead{
\colhead{Model Name} & \colhead{Learning Rate} & \colhead{Momentum} & \colhead{L2 Regularization ($\lambda$)} & \colhead{Batch Size} & \colhead{Dropout Rate}
}
\startdata
    \hline
    \hline
    Low-z Sim. Trained & $5\times10^{-7}$ & 0.99 & $10^{-4}$ & 16 & $7\times10^{-4}$\\
    Mid-z Sim. Trained & $5\times10^{-7}$ & 0.99 & $10^{-4}$ & 16 & $7\times10^{-4}$\\
    High-z Sim. Trained & $5\times10^{-7}$ & 0.99 & $10^{-4}$ & 16 & $4\times10^{-4}$\\
    \hline
    Low-z Final & $5\times10^{-8}$ & 0.99 & $10^{-4}$ & 16 & $4\times10^{-4}$\\
    Mid-z Final & $5\times10^{-8}$ & 0.99 & $10^{-4}$ & 16 & $2\times10^{-4}$\\
    High-z Final & $5\times10^{-6}$ & 0.99 & $10^{-4}$ & 16 & $2\times10^{-4}$\\
\enddata
\end{deluxetable*}

\section{Identifying Issues with our Light Profile Fitting Pipeline}\label{ap:sec:galfit_failures}

We described in \S \ref{sec:galfitting} a semi-automated pipeline that we used to determine the structural parameters for $\sim60,000$ galaxies using light-profile fitting. After performing this analysis, we visually inspected the fits of a randomly selected sub-sample of $\sim300$ galaxies from each redshift bin, to assess the quality of the fit. 

The process of visually inspecting $\sim 900$ galaxies helped us to identify three failure modes of our fitting pipeline:-

\begin{enumerate}
    \item for some galaxies, GALFIT assigned an extremely small axis ratio to one of the components leading to an unphysical lopsided bulge/disk
    \item for some galaxies, the centroids of the two fitted components were too far away from each other
    \item some galaxies had residuals that were too large
\end{enumerate}

\begin{figure*}[htb]
    \centering
    \includegraphics[width = 0.9\textwidth]{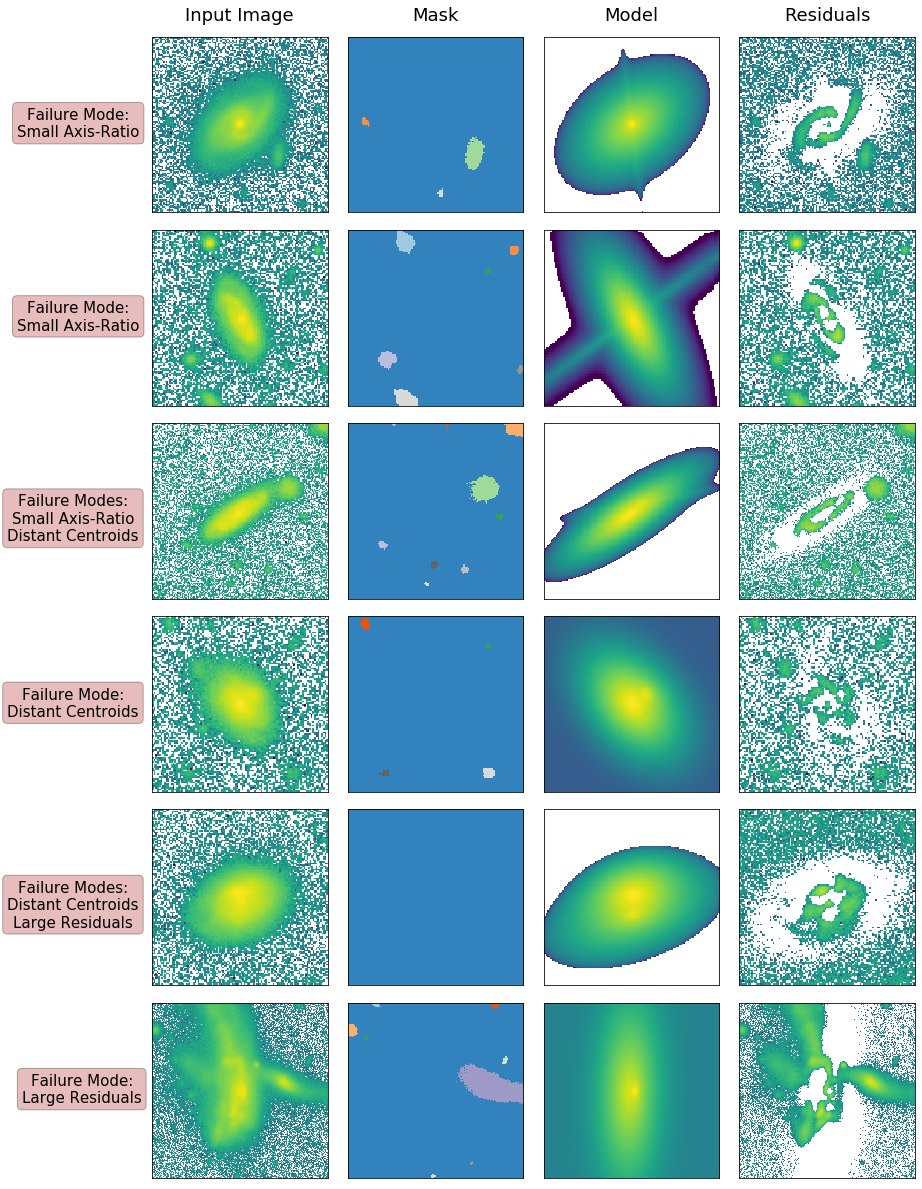}
    \caption{The different failure modes of the semi-automated light profile fitting code described in \S \ref{sec:galfitting}. From left to right, we show the input image, the  mask generated by Source Extractor, the model generated by GALFIT, and the residuals.}
    \label{fig:galfit_fails}
\end{figure*}

Note that for some galaxies, multiple failure modes were applicable. Examples of each of these failure modes are shown in Figure \ref{fig:galfit_fails}. We use a combination of different cuts on the fitted dataset to get rid of these failure modes -- see \S \ref{sec:galfitting} for an extended discussion. 

\section{Additional Two-Dimensional Residual \& Uncertainty Plots} \label{ap:sec:2d_residuals}

\begin{figure*}[htb]
    \centering
    \includegraphics[width = 0.9\textwidth]{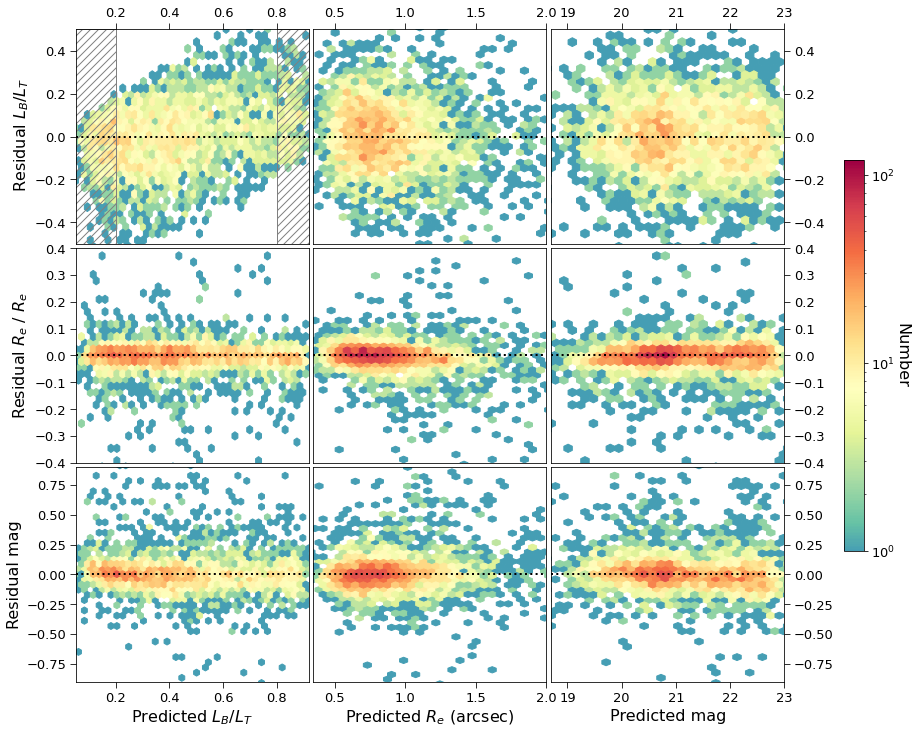}
    \caption{Residuals of the output parameters (difference between \gampen{} and GALFIT predictions) plotted against the values predicted by \gampen{} for all galaxies in the mid-z test set. See \S\,\ref{sec:gampen_v_galfit} for details.}
    \label{fig:2d_res_mid_z}
\end{figure*}

\begin{figure*}[htb]
    \centering
    \includegraphics[width = 0.9\textwidth]{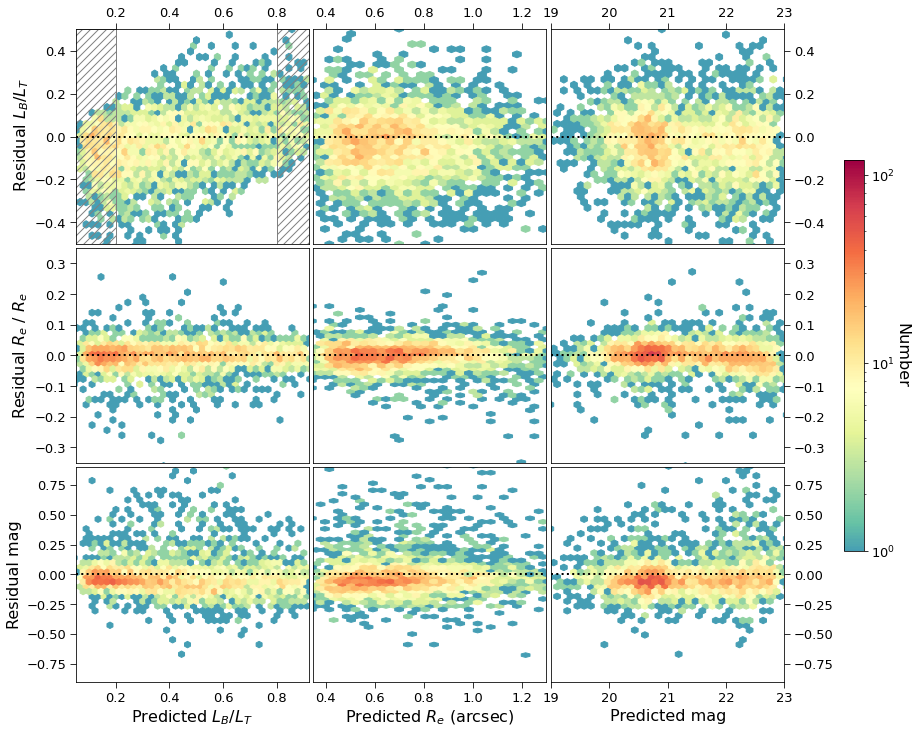}
    \caption{Residuals of the output parameters (difference between \gampen{} and GALFIT predictions) plotted against the values predicted by \gampen{} for all galaxies in the high-z test set. See \S\,\ref{sec:gampen_v_galfit} for details.}
    \label{fig:2d_res_high_z}
\end{figure*}

Figures \ref{fig:2d_res_mid_z} and \ref{fig:2d_res_high_z} show the distribution of residuals  (difference between \gampen{} and GALFIT predictions) for the mid- and high-z bins. 

\begin{figure*}[htb]
    \centering
    \includegraphics[width = 0.9\textwidth]{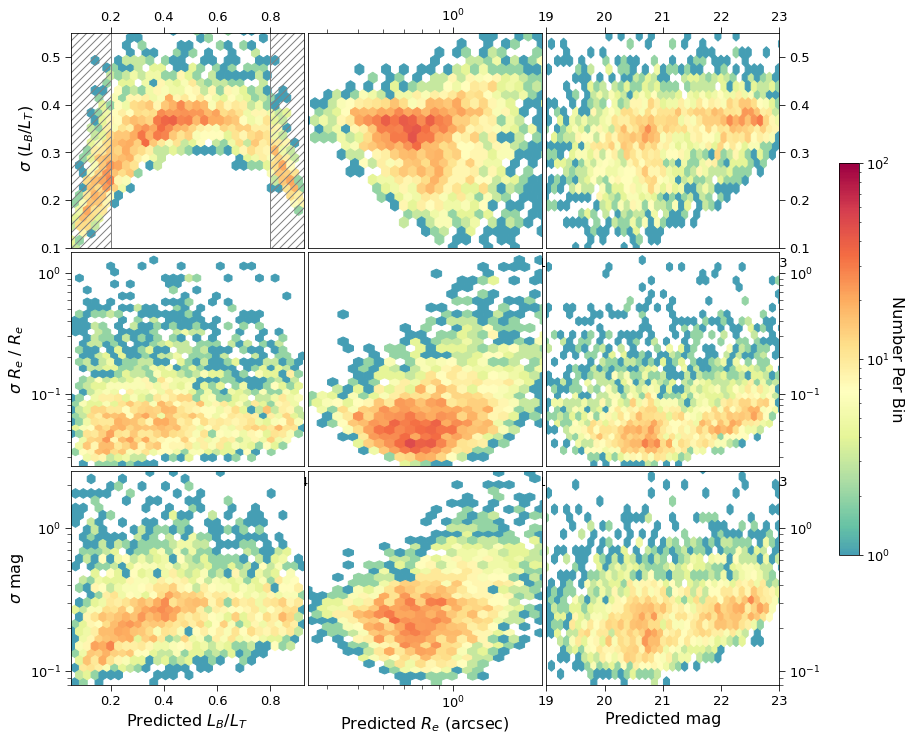}
    \caption{Uncertainties predicted by \gampen{} for each parameter plotted against the predicted values for the mid-z test set. The $\sigma$ for each parameter is defined as the width of the $68.27\%$ confidence interval. The line-shaded region in the top-left panel shows the region where we recommend transforming quantitative $L_B/L_T$ predictions to qualitative labels. See \S \ref{sec:uncertainties} for details.}
    \label{fig:2d_uncer_mid_z}
\end{figure*}

\begin{figure*}[htb]
    \centering
    \includegraphics[width = 0.9\textwidth]{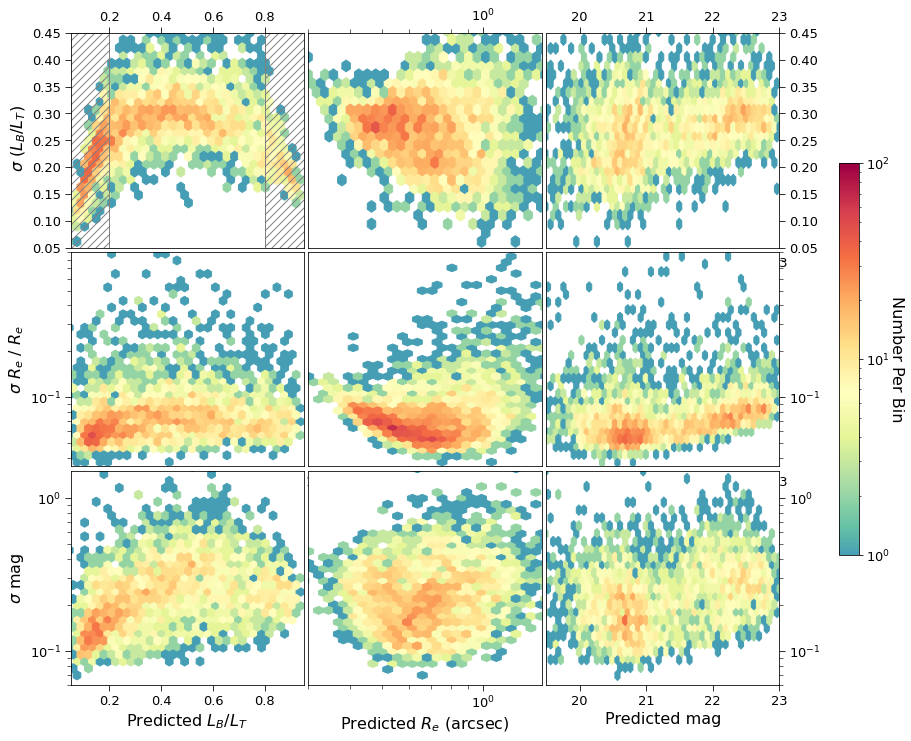}
    \caption{Uncertainties predicted by \gampen{} for each parameter plotted against the predicted values for the high-z test set. The $\sigma$ for each parameter is defined as the width of the $68.27\%$ confidence interval. The line-shaded region in the top-left panel shows the region where we recommend transforming quantitative $L_B/L_T$ predictions to qualitative labels. See \S \ref{sec:uncertainties} for details.}
    \label{fig:2d_uncer_high_z}
\end{figure*}

Figures \ref{fig:2d_uncer_mid_z} and \ref{fig:2d_uncer_high_z} show the uncertainties predicted by \gampen{} for each parameter plotted against the predicted values for the mid- and high-z bins.

\section{Comparing \gampen{} and GALFIT's performance on smaller simulated galaxies} \label{ap:sec:gapemn_v_galfit}

\begin{figure*}[htb]
    \centering
    \includegraphics[width = 0.9\textwidth]{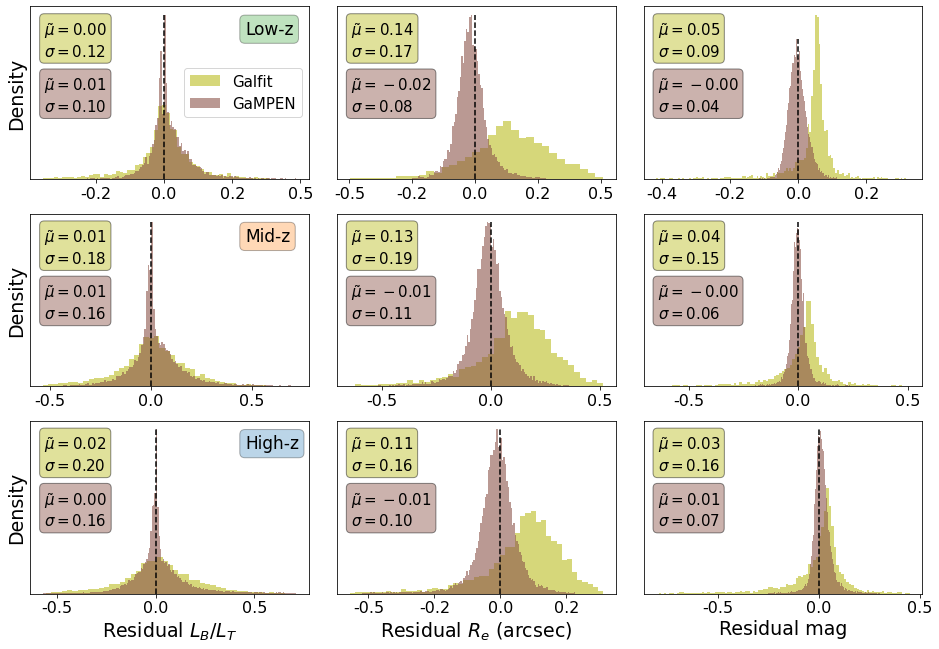}
    \caption{Distribution of residuals for $\sim5000$ simulated galaxies in each redshift bin with $R_e \leq 2\arcsec$.  These galaxies were selected randomly from the simulation testing set. The top, middle, and bottom rows show the results for the low-, mid-, and high-z bins respectively. The boxes in the top-left corner of each panel show the median ($\tilde{\mu}$) and standard deviation ($\sigma$) of each residual distribution. The dashed black vertical line marks $x=0$. }
    \label{fig:gampen_v_galfit}
\end{figure*}

As shown in \S \ref{sec:gampen_v_galfit}, GaMPEN and GALFIT systematically disagree more for galaxies with smaller sizes. To ascertain their relative performance, specifically for smaller galaxies, we ran our GALFIT pipeline (described in \S \ref{sec:galfitting}) on $\sim5000$ simulated galaxies from each redshift bin with $R_e \leq 2\arcsec$. These galaxies were chosen randomly from the testing set of \gampen{} -- thus, none of them were used to train \gampen{}. Thereafter, we compared the results of this fitting procedure to the predictions made by \gampen{} on the same galaxies. The residuals obtained for both \gampen{} and GALFIT are shown in Figure \ref{fig:gampen_v_galfit}.

The typical error for each of the parameters is given by $\tilde{\mu}\pm\sigma$, where $\tilde{\mu}$ and $\sigma$ are the median and standard deviation of the residual distribution respectively. As shown in Figure \ref{fig:gampen_v_galfit}, \gampen{} outperforms GALFIT for all three parameters across all redshift bins. This provides preliminary evidence that \gampen{}'s predictions on the smaller galaxies referred to in \S \ref{sec:gampen_v_galfit} are more accurate than those obtained using GALFIT.

\software{PyTorch \citep{pytorch},
          Ignite \citep{ignite},
          MLFlow \citep{mlflow},
          Numpy \citep{numpy},
          Scipy \citep{Virtanen2020SciPyPython},
          Astropy \citep{astropy_1,astropy_2},
          Pandas \citep{pandas},
          Scikit-learn \citep{scikitlearn},
          Matplotlib \citep{matplotlib},
          }

\bibliographystyle{aasjournal}
\bibliography{references}

\end{CJK*}
\end{document}